\documentclass[aps,prd,10pt,showpacs,amsmath,showkeys,twocolumn,floatfix,amssymb, preprintnumbers, nofootinbib, superscriptaddress]{revtex4}
\usepackage{graphicx}
\usepackage{comment}
\usepackage[usenames]{color}
\usepackage{bm}
\usepackage{ifpdf}
\usepackage{floatrow}
\usepackage{makecell}
\usepackage{caption}
 \usepackage{subcaption}

\usepackage[normalem]{ulem}
\usepackage[dvipsnames]{xcolor}
\usepackage[utf8]{inputenc}
\usepackage{hyperref}
\hypersetup{
  pdfnewwindow=true,      
  colorlinks=true,        
  linkcolor=PineGreen,    
  citecolor=PineGreen,    
  filecolor=PineGreen,    
  urlcolor=PineGreen      
}
\newcommand{\be}{\begin{eqnarray}}
\newcommand{\ee}{\end{eqnarray}}

\begin{document}

\title{Tunneling time in  \texorpdfstring{$\mathcal{P}\mathcal{T}$}{PT}-symmetric systems}

\author{Peng~Guo}
\email{pguo@csub.edu}
\affiliation{Department of Physics and Engineering,  California State University, Bakersfield, CA 93311, USA}

\author{Vladimir~Gasparian}
\email{vgasparyan@csub.edu}
\affiliation{Department of Physics and Engineering,  California State University, Bakersfield, CA 93311, USA}

\author{Esther~J\'odar}
\email{esther.jferrandez@upct.es}
\affiliation{Departamento de F\'isica Aplicada,  Universidad Polit\'ecnica de Cartagena, E-30202 Murcia, Spain}

\author{Christopher~Wisehart }
\email{cwisehart@csub.edu}
\affiliation{Department of Physics and Engineering,  California State University, Bakersfield, CA 93311, USA}

\date{\today}

\begin{abstract}
In  the present work  we propose a  generalization of tunneling time in  parity and time ($\mathcal{P}\mathcal{T}$)-symmetric systems. The properties of  tunneling time in  $\mathcal{P}\mathcal{T}$-symmetric systems  are studied   with a simple contact interactions periodic finite size diatomic $\mathcal{P}\mathcal{T}$-symmetric  model. The physical meaning of negative tunneling time in  $\mathcal{P}\mathcal{T}$-symmetric systems and its relation to spectral singularities is discussed. 
\end{abstract}

\maketitle

\section{Introduction}\label{sec:intro}

In the standard (Hermitian) quantum mechanics the calculation of the time interval during which
a particle interacts with a barrier of arbitrary shape is not new and  has raised large
interest lately.  A great variety of theoretical and
experimental work on this topic has been carried out during the last two decades.  
This particular interest is due
to the introduction of non-Hermitian elements into the Hamiltonian enabling the solution of
specific problems where, for example, a complex index of refraction  is used (optics) or where complex potentials are introduced far away from the interaction region
of the particles \cite{moiseyev_2011}.
The tunneling time problem was studied both
theoretically and experimentally (see e.g., Refs.~\cite{RevModPhys.66.217,GASPARIAN2000513,fayer2001elements}  
and references therein), especially in nanostructures 
or in mesoscopic systems smaller than 10 $nm$. In
these systems the tunnelling time will eventually
play an important role in determining transport properties, for example in the frequency-dependent conductivity response of mesoscopic conductors \cite{Buttiker1994} and
in the phenomenon of an adiabatic charge transport
\cite{PhysRevB.58.R10135,PhysRevLett.82.608}. Several authors have studied the problem of tunneling time in passive scattering systems using
a number of completely different approaches, including the oscillatory amplitude of the incident amplitude \cite{LEAVENS19871101,MButtiker_1985,Buttiker5390141,PhysRevB.45.1742}, the time-modulated barrier \cite{PhysRevLett.49.1739}  and as well as wave packet approach, see, e.g. Refs.~\cite{LandauerNature1993,PhysRevA.60.1811,MUGA199224}.  In most approaches to the tunnelling time problem more than one tunneling-time
component is involved, regardless of whether we deal with the so-called B{\"u}ttiker–Landauer \cite{PhysRevB.27.6178}, 
the Feynman path-integral approach \cite{PhysRevA.36.4604} or complex characteristic
times ~\cite{PhysRevB.47.2038,PhysRevB.51.6743,PhysRevA.54.4022}. Furthermore, this
does not seem to be a peculiarity of quantum mechanical waves, but a general result and valid not
only for electrons but also for any waves (sound and electromagnetic), when their
propagation through a medium is described by a differential equation of second order. 
For electromagnetic waves, 
in Ref.~\cite{PhysRevLett.75.2312}  in the Faraday
rotation scheme, it was shown that the characteristic time associated with the interaction of the classical electromagnetic wave with a finite region is a complex quantity. As a consequence, the emerging electromagnetic wave is elliptically polarized, and the major axis of the ellipse is rotated relative to the original direction of polarization. In addition, one of the components of the complex time is closely related not only to the total density of states (DOS), but also to its decomposition into partial DOS \cite{fayer2001elements,PhysRevB.27.6178}.  Furthermore, the two 
components are not entirely independent quantities and connected by Kramers–Kronig relations \cite{Gasparian1999}. In Ref.~\cite{PhysRevLett.78.851} experimentally investigated the
tunneling times associated with frustrated total internal
reflection of light. It is shown that the two characteristic times correspond, respectively, to the spatial and
angular shifts of the beam. 
Despite the progress which has been made towards understanding the longstanding problem of tunneling time in traditional quantum mechanics, as far as we know, such discussions are rather scarce in non-Hermitian systems  (see, e.g., Refs. \cite{MUGA2004357,Hasan2020,JIAN2020125815}). 

Without pretending to give an exhaustive overview of the possible applications of non-Hermitian systems as an ideal platform for exploring the functionality of new possible devices, we briefly present some achievements in condensed matter physics \cite{PhysRevLett.103.030402}, electronic circuits \cite{PhysRevA.84.040101}, coupled mechanical oscillators \cite{PhysRevA.88.062111,doi:10.1119/1.4789549,PhysRevA.90.022114}, mesoscopic superconducting wires \cite{PhysRevLett.109.150405} and photonic applications \cite{Feng2017}.

The most studies in non-Hermitian systems, both theoretical and experimental, have been carried out in 
parity-time ($\mathcal{P}\mathcal{T}$) symmetric systems, supporting real spectrum of eigenvalues. In such a system great attention was paid to the optical setups or theoretical models with the gain (through optical or electrical
pumping or nonlinear interactions) and loss (due to absorption or
radiation) where the properties of the $\mathcal{P}\mathcal{T}$-symmetric  system, including 
unidirectional invisibility \cite{PhysRevLett.106.213901,Feng2013,PhysRevA.87.012103} and double refraction \cite{PhysRevLett.100.103904} can be measured directly or calculated analytically. 
In Ref.~\cite{PhysRevLett.110.223902}  there have been investigated, both theoretically
and experimentally, the instabilities and the possibility of establishing localized complex defect modes with
spectra lying within the allowed band. An interesting feature of resonance, as in discrete, as well as in scattering spectra in a $\mathcal{P}\mathcal{T}$-symmetric open quantum system
was studied in Ref.  \cite{PhysRevA.92.022125},
using the tight-binding model. 

After this brief enumeration of the intriguing features of $\mathcal{P}\mathcal{T}$-symmetric systems, the need for further research on the problem of tunneling time in $\mathcal{P}\mathcal{T}$-symmetric systems becomes apparent. Moreover,  as shown in Ref.~\cite{PhysRevResearch.4.023083}, the absorbing part
of    Green's function of non-Hermitian systems is no longer related to the density of states, and it is
generally a complex function. Only in special
cases, such as $\mathcal{P}\mathcal{T}$-symmetric systems, the absorbing part of
  Green's function can remain real, although a positive-definite norm is not guaranteed. The purpose of the present work goes in this direction, in
the sense that  the concept of generalized tunneling time in $\mathcal{P}\mathcal{T}$-symmetric systems is proposed. The properties of  tunneling time in  $\mathcal{P}\mathcal{T}$-symmetric systems is studied with a simple contact interactions periodic finite size diatomic $\mathcal{P}\mathcal{T}$-symmetric  model. We stress, that unlike the positively definite tunneling time $\tau_1$ (see below, Eq. (\ref{tau1})) in real
potential scattering theory, the $\tau_1$ in $\mathcal{P}\mathcal{T}$-symmetric systems can be either positively or negatively valued. The
value of $\tau_1$  turning negative is closely related to the motion of pole singularities of scattering amplitudes in complex energy plane. The physical interpretation of negative valued $\tau_1$ is also discussed.

The paper is organized as follows.   The generalized tunneling time for $\mathcal{P}\mathcal{T}$-symmetric systems is introduced and  defined in Sec.~\ref{timePTdef}. A simple model  and its analytic solutions are summarized in Sec.~\ref{PTmodel}. The averaged tunneling time per unit cell and its limit to an infinite periodic system is  discussed in Sec.~\ref{avgtime}. The impact of spectral singularities and  numerics are shown in Sec.~\ref{sec:spectralsingularity}. Followed by the discussions and summary in   Sec.~\ref{summary}.

\section{Generalized tunneling time in   \texorpdfstring{$\mathcal{P}\mathcal{T}$}{PT}-symmetric systems}\label{timePTdef}  
 In the real potential scattering theory, the concept of tunneling or delay time for a particle tunneling through potential barriers is conventionally defined through integrated density of state, which is proportional to the imaginary part of full Green's function of systems,   see e.g. Refs.~\cite{PhysRevB.47.2038,PhysRevB.51.6743,PhysRevA.54.4022},
 \begin{equation}
\tau_1 (E) = -Im \left [  \int_{- l}^l d x \langle x | \hat{G} (E) |  x \rangle  \right ]  \propto  \int_{- l}^l d x  |\langle x | \Psi_E \rangle |^2, \label{tau1}
 \end{equation} 
  where $l$ is the  half length of potential barrier.  The $\hat{G} (E)  =  \frac{1}{ E- \hat{H} } $ is the Green's function operator of system.  It is related to eigenstate of system   by  the spectral representational of Green's function,
   \begin{equation}
  \hat{G} (E)   = \sum_{i} \frac{ |  \Psi_{E_i} \rangle \langle \Psi_{E_i } |}{E- E_i},
 \end{equation} 
 where  eigenstate $|  \Psi_{E_i} \rangle$ satisfies Schr\"odinger equation,
 \begin{equation}
\hat{H} |  \Psi_{E_i} \rangle = E_i |  \Psi_{E_i} \rangle.
 \end{equation}

  In the case of complex potential scattering theory, now one is facing the challenge of how the concept of tunneling time should be  generalized and defined properly.  One of  the key elements of developing the concept of tunneling or delay time in a real potential scattering theory    is counting the probability that a particle  spends  inside of a barrier, see  e.g. Refs.~\cite{PhysRev.98.145,PhysRev.118.349,goldberger2004collision}. However, in complex potential scattering theory, the norm of states is no longer conserved,  so the probability interpretation of tunneling time becomes problematic. In addition, the spectral representation of Green's function now depends on the eigenstates of both $\hat{H} $ and its adjoint $\hat{H}^\dag $,
   \begin{equation}
  \hat{G} (E)   = \sum_{i} \frac{ |  \Psi_{E_i} \rangle \langle \widetilde{\Psi}_{E_i } |}{E- E_i},
 \end{equation}  
  where
   \begin{equation}
\hat{H}^\dag |  \widetilde{\Psi}_{E_i} \rangle = E_i |  \widetilde{\Psi}_{E_i} \rangle.
 \end{equation}
  The biorthogonal and normalization relations can only be established  by eigenstates of  dual systems together \cite{FESHBACH1985398,MUGA2004357,Brody_2013},
    \begin{equation}
   \sum_{i}   |  \Psi_{E_i} \rangle \langle \widetilde{\Psi}_{E_i } |  = \mathbb{I}.
 \end{equation}  
 Hence, the discontinuity of Green's function crossing the branch cut in complex $E$-plane is in general  a complex function,  and it is no longer equivalent to the imaginary part of Green's function, see  Ref.~\cite{PhysRevResearch.4.023083}.   Its relevance to the conventional definition of density of  states is also lost,
    \begin{equation}
Disc_E    \langle  x |  \hat{G} (E) |  x \rangle   \propto    \langle  x |  \Psi_{E} \rangle    \langle \widetilde{\Psi}_{E } |  x \rangle \neq  |\langle x | \Psi_E \rangle |^2, 
 \end{equation} 
 where 
    \begin{equation}
Disc_E      \hat{G} (E) = \frac{ \hat{G} (E+ i 0) -  \hat{G} (E - i 0) }{2 i }.
 \end{equation}

Fortunately,  in  a $\mathcal{P}\mathcal{T}$-symmetric  system, as shown in Ref.~\cite{PhysRevResearch.4.023083}, because of symmetry constraints,  the discontinuity of Green's function is still a real function, hence is identical to the imaginary part of Green's function, $Disc_E      \hat{G} (E) = Im  \hat{G} (E) $. However, as   the consequence of  norm violation with a complex potential,   positivity of the imaginary part of Green's function is not guaranteed. In this work, the tunneling time through $\mathcal{P}\mathcal{T}$-symmetric  barriers is still   defined by an integrated Green's function, hence most of formalism that are developed for real potential scattering theory can still apply directly to $\mathcal{P}\mathcal{T}$-symmetric  system. The  imaginary part of Green's function is now referred as generalized density of  states of a $\mathcal{P}\mathcal{T}$-symmetric  system. Following the definition in Refs.~\cite{PhysRevB.47.2038,PhysRevB.51.6743,PhysRevA.54.4022},  two components of the traversal time $\tau_E$ are introduced by
 \begin{equation}
 \tau_E = \tau_2 (E) + i \tau_1 (E) = - \int_{ -  l}^{ l } d x \langle x | \hat{G} (E) | x \rangle , \label{taudef}
 \end{equation}
where $\tau_1$ and $\tau_2$ may be interpreted as two components of  generalized concept of   B\"uttiker-Landauer  tunneling time  that are connected with  generalized density of states and the Landauer resistance in a  $\mathcal{P}\mathcal{T}$-symmetric  system.  The positivity and negativity of generalized tunneling time $\tau_1$   simply  reflects the nature of potential barriers that either tend to keep a particle in or force it out. The   negative value portion  of $\tau_1$ is thus physically inaccessible  and hence behave similar to a forbidden gap in a periodic system.

 As shown in Refs.~\cite{PhysRevB.51.6743,PhysRevA.54.4022},   the integrated Green's function is related to the transmission and reflection amplitudes by   
  \begin{align}
&  - \int_{ -  l}^{ l } d x \langle x | \hat{G} (E) | x \rangle   \nonumber \\
& = \frac{d}{d E} \ln \left [t (k)e^{i k 2 l} \right ]+ \frac{ r^{(L)} (k) + r^{(R)} (k) }{4 E} e^{i k 2 l}, \label{integGtr}
 \end{align}
  where  $t(k)$ and $r^{(L/R)}(k)$ are the transmission and left/right reflection amplitudes respectively, and $k = \sqrt{2 m E}$ is the momentum of particle.  In practice,  it is more convenient to compute the tunneling time  directly through the transmission and reflection amplitudes.

  \begin{figure*}
 \centering
 \begin{subfigure}[b]{0.49\textwidth}
\includegraphics[width=0.99\textwidth]{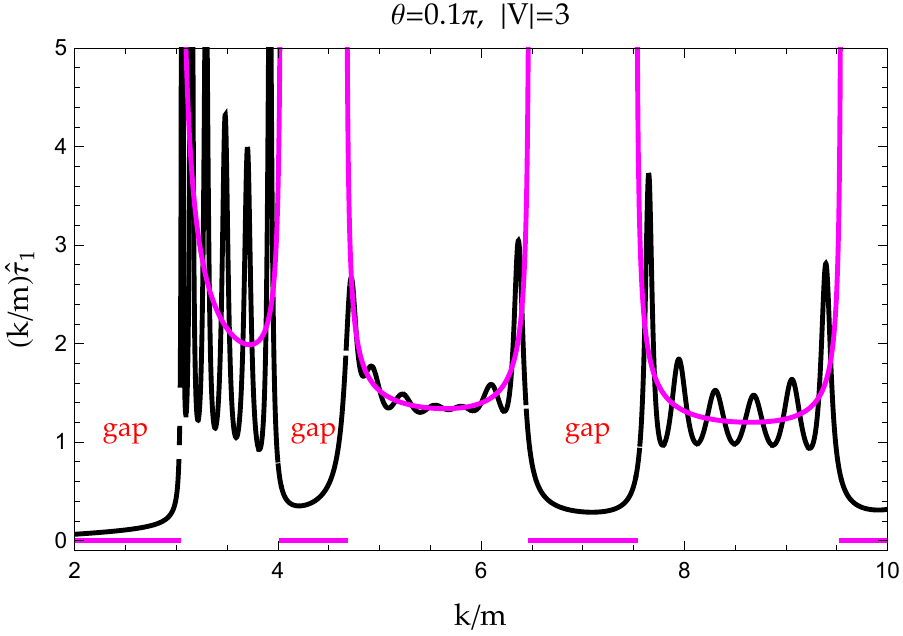}
\caption{     }\label{tau1plot1}
\end{subfigure} 
\begin{subfigure}[b]{0.49\textwidth}
\includegraphics[width=0.99\textwidth]{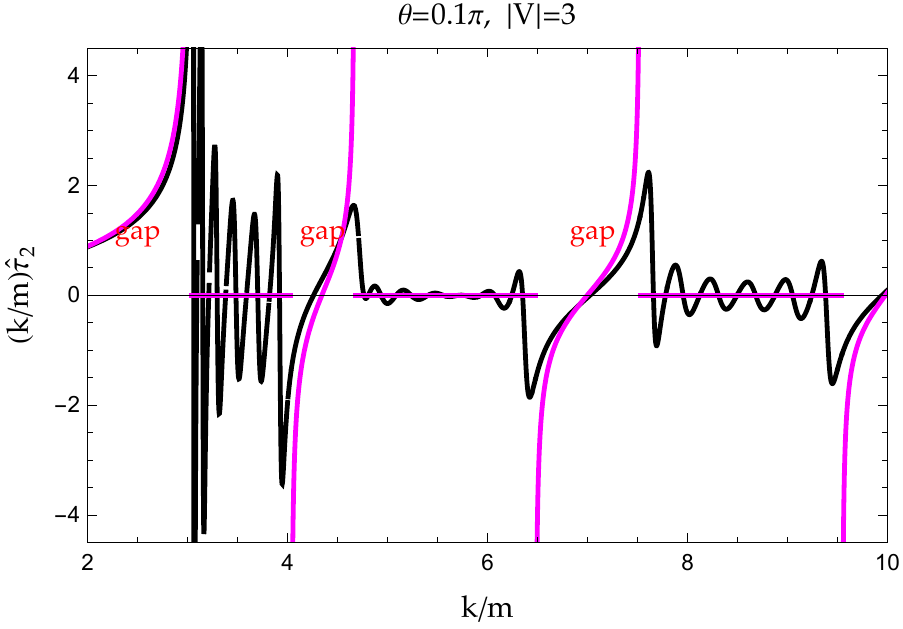}
\caption{     }\label{tau2plot1}
\end{subfigure}
\begin{subfigure}[b]{0.49\textwidth}
\includegraphics[width=0.99\textwidth]{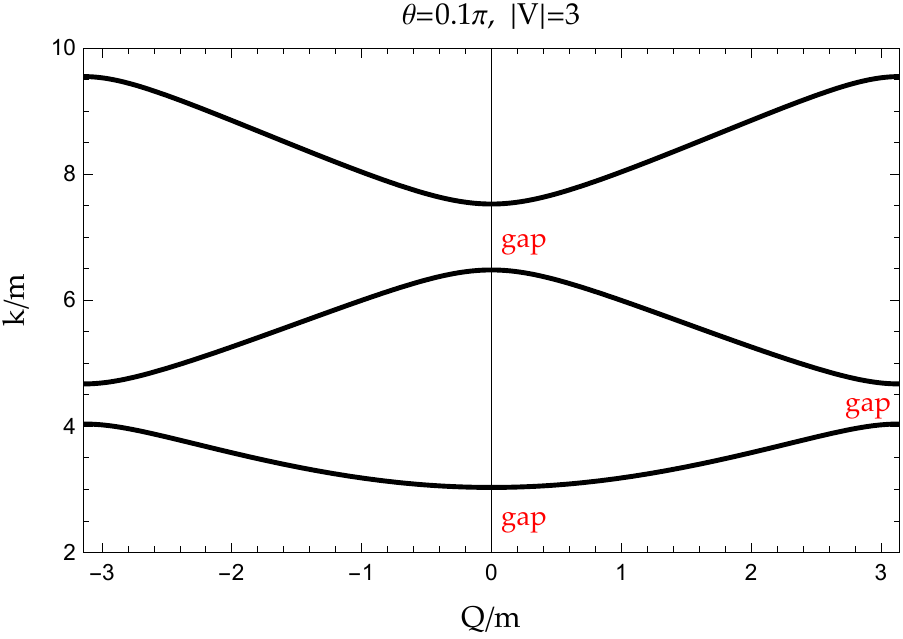}
\caption{    }\label{bandplot1}
\end{subfigure}
\caption{   (a) and (b) Comparison of   $\frac{k}{m} \widehat{\tau}$ with $N=3$ (solid black)  together with        $i \frac{k}{m} \frac{d Q}{d E} $     (solid purple/light grey)  vs. $k/m$. More specifically (a) $\frac{k}{m} \widehat{\tau}_1$  (solid black)  together with $\frac{k}{m} \frac{d Re[Q]}{d E} $  (solid purple/light grey); (b) $\frac{k}{m} \widehat{\tau}_2$   (solid black)  together with $ \frac{k}{m} \frac{d Im [Q ]}{d E} $ (solid purple/light grey); (c) The corresponding band structure plot  in unbroken   $\mathcal{P}\mathcal{T}$-symmetric phase.  The parameters are taken as:   $\theta =0.1 \pi$, $|V| =3 $, $m L=1$ and $m a =0.2$, where $|V|$ is dimensionless. }\label{tauplot1}
\end{figure*}

  \begin{figure*}
 \centering
 \begin{subfigure}[b]{0.49\textwidth}
\includegraphics[width=0.99\textwidth]{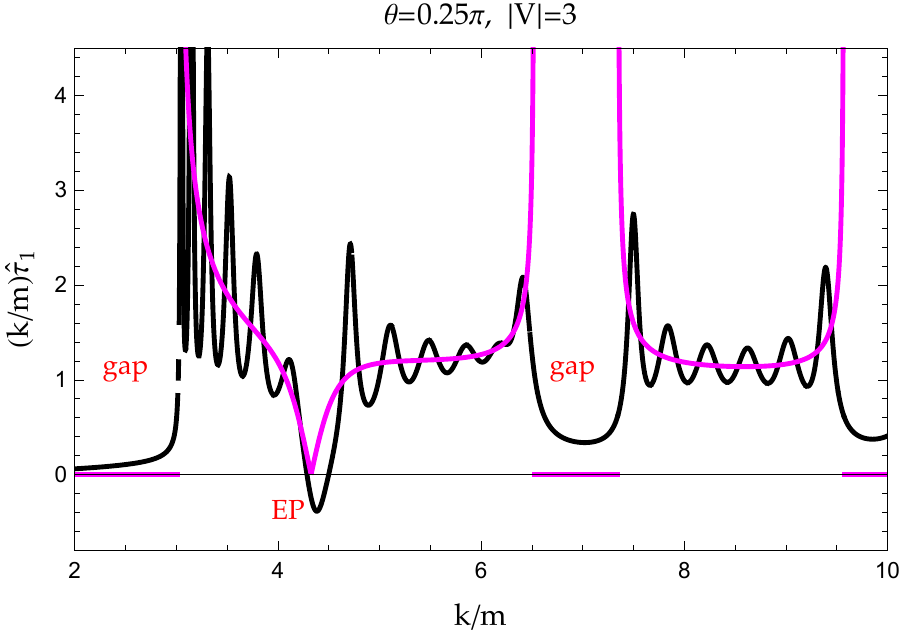}
\caption{  }\label{tau1plot2}
\end{subfigure} 
\begin{subfigure}[b]{0.49\textwidth}
\includegraphics[width=0.99\textwidth]{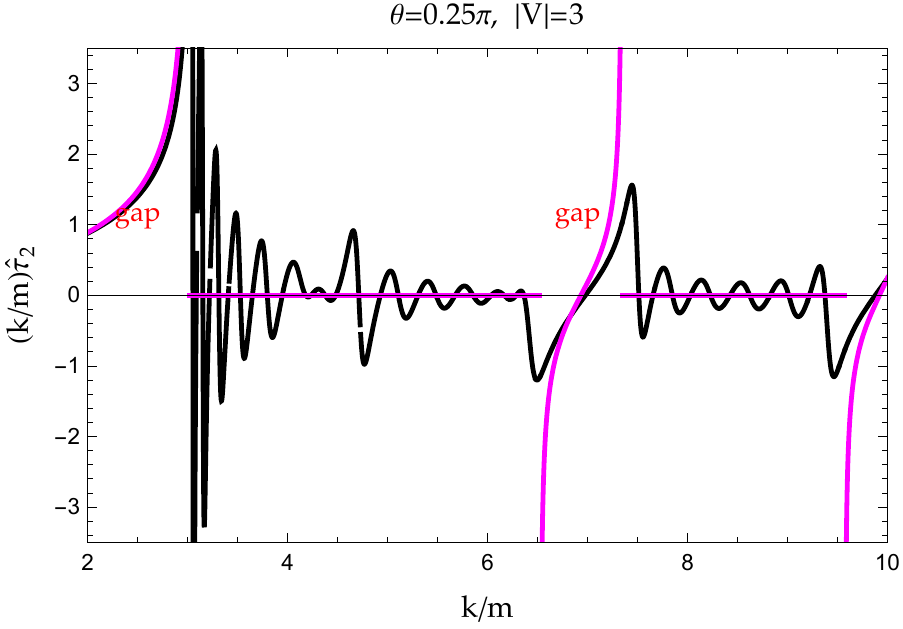}
\caption{      }\label{tau2plot2}
\end{subfigure}
\begin{subfigure}[b]{0.49\textwidth}
\includegraphics[width=0.99\textwidth]{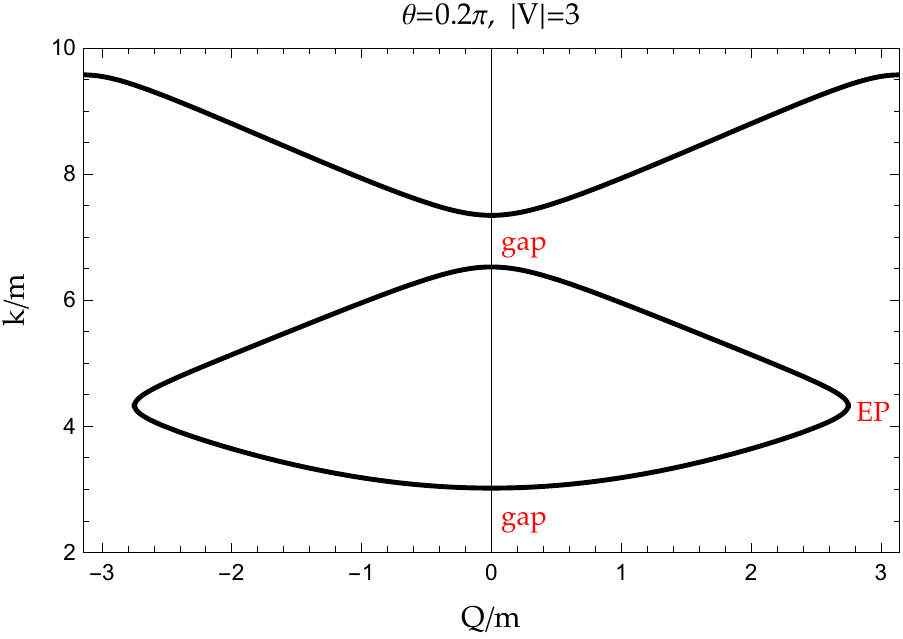}
\caption{  }\label{bandplot2}
\end{subfigure}
\caption{   (a) and (b) Comparison of   $\frac{k}{m} \widehat{\tau}$ with $N=3$ (solid black)  together with     $i \frac{ k}{m} \frac{d Q}{d E} $    (solid purple/light grey)  vs. $k/m$. More specifically (a)   $\frac{ k}{m} \widehat{\tau}_1$   (solid black)  together with $\frac{ k}{m} \frac{d Re[Q]}{d E} $  (solid purple/light grey);  (b) $\frac{ k}{m} \widehat{\tau}_2$  (solid black)  together with $ \frac{ k}{m} \frac{d Im [Q ]}{d E} $  (solid purple/light grey); (c)  The corresponding band structure plot  in broken   $\mathcal{P}\mathcal{T}$-symmetric phase. The parameters are taken as:   $\theta =0.2 \pi$, $|V| =3 $, $m L=1$ and $m a =0.2$, where $|V|$ is dimensionless. }\label{tauplot2}
\end{figure*}

\section{A simple contact interactions \texorpdfstring{$\mathcal{P}\mathcal{T}$}{PT}-symmetric impurities model}\label{PTmodel}

 In this section,  a simple contact interactions $\mathcal{P}\mathcal{T}$-symmetric model is adopted to investigate  the properties of tunneling time   through  layers of $\mathcal{P}\mathcal{T}$-symmetric barriers.   Each single cell of barrier  is composed of two complex components: one absorbing component with loss and another emissive component with gain. Two components are placed symmetrically on    two sides of the cell along center, the gain and loss in each single cell are balanced. The potential barrier in a single cell hence is $\mathcal{P}\mathcal{T}$-symmetric: 
  \begin{equation}
  V(-x) = V^*(x).
  \end{equation}
  The Hamiltonian of  barriers is thus given by
   \begin{equation}
\hat{H} = - \frac{1}{2 m} \frac{d^2}{d x^2} + \sum_{n= - N}^N V(x- n L) 
 \end{equation}
  where    $2 N+1$   is the total number of cells  and $L$ represents the length of a single cell. The total length of barriers is $2 l  = (2N+1) L$. 
A  simple diatomic  $\mathcal{P}\mathcal{T}$-symmetric contact interaction  potential  that represents two complex conjugate impurities is used:
  \begin{equation}
  V(x) = V \delta(x -a) + V^* \delta(x+a), \ \ \ \ V = |V| e^{i \theta}. 
  \end{equation}
  Two  complex conjugate impurities  are placed on two sides of a cell's center with equal distance $a$. One is absorbing with loss and another is emissive with equal amount gain. The physical realization of such a  $\mathcal{P}\mathcal{T}$-symmetric model may be accomplished in a planar slab waveguide as discussed in Ref.~\cite{Ruschhaupt_2005}.
  
The simple contact interaction  $\mathcal{P}\mathcal{T}$-symmetric impurities model adopted in this work may be considered as  a special case of periodic   diatomic contact interactions model. It turns out the analytic form of scattering solutions can be found in a highly non-trivial way, see e.g. characteristics determinant approach in Refs.~\cite{GASPARIAN1988201,GASPARIAN199772,Aronov_1991}.    In  what follows, we will simply summarize the results of scattering solutions. A brief introduction about the general scattering theory, some useful relations and characteristics determinant approach  are provided in Appendix~\ref{scattgen}.

The transmission and left/right reflection amplitudes are given by
 \begin{align}
 t(k)   &= \frac{ \sec (Q(2N+1)  L ) e^{ - i k (2N+1) L} }{   1+ i  Im \left [ \frac{e^{- i k L} }{t_0(k)}    \right ] \frac{\tan  (Q (2N+1) L)}{\sin (Q L)}    },  \nonumber \\
 \frac{ r^{(L/R)} (k) }{t(k) } & = \left [  \frac{ r^{(L/R)}_0 (k) }{t_0(k)}  \right ] \frac{\sin (Q (2N+1) L)}{\sin (QL) }     , \label{tandrexpress}
\end{align}
where  $t_0 (k)$  and $ r^{(L/R)}_0 (k)$ are   transmission and reflection amplitudes by a single cell ($N=0$),
\begin{align}
& \frac{1 }{t_0(k)}  =  1+ 2 \frac{ i m  | V|  }{k  } \cos \theta   + 2 i  \left  ( \frac{  m|V| }{k} \right  )^2 \sin ( k 2 a) e^{  i  k 2a    }    , \nonumber \\
 & \frac{ r^{(L/R)}_0 (k) }{t_0(k)}   =  - 2 \frac{i m |V|}{k}    \left [   \cos (    k2 a \mp \theta) +  \frac{ m |V|}{k}   \sin (k 2a)      \right ] . \label{t0r0LReqs}
\end{align}
The $Q$ plays the role of crystal-momentum for a periodic lattice and is related to $k$ by
\begin{equation}
 \cos (Q L ) = Re \left [ \frac{e^{- i k L} }{t_0(k)}   \right ]   .  \label{cosQLRet0}
\end{equation}
It is quite remarkable to see that the transmission and reflection amplitudes for a periodic many  scatters system are all related to single cell scattering amplitudes in a very compact fashion.  If $Q$ is treated as a free parameter that represents the collective mode of entire lattice of all impurities, thus the short-range interaction dynamics that is described by single cell scattering amplitudes and long-range physics of collective mode are totally factorized. The factorization  of  short-range dynamics and long-range correlation for an infinite long periodic lattice or in a periodic finite box has been a well-known fact in both condensed matter physics and nuclear/hadron physics.  When the interaction range is much smaller than the size of a cell or a trap,  the quantization conditions are given by a compact formula, in which  two components: (1) the short-range particles dynamics and (2) long-range geometric effect due to the  periodic box or trap  are   well separated. The compact form is known as Korringa–Kohn–Rostoker  (KKR) method  \cite{KORRINGA1947392,PhysRev.94.1111} in condensed matter physics, L\"uscher formula  \cite{Luscher:1990ux}  in LCQD  and  Busch-Englert-Rza\.zewski-Wilkens (BERW) formula \cite{Busch98} in a harmonic oscillator trap in nuclear physics community. Other related useful discussions can be found in  e.g. Refs.\cite{Guo_2022_JPG,PhysRevD.103.094520,Guo_2022_JPA,PhysRevC.103.064611},   also see the discussion in Appendix~\ref{finitevolumesol}.  The emergence of periodic dynamics in a finite size system may be best demonstrated by Characteristic determinant approach that was developed in Refs.~\cite{GASPARIAN1988201,GASPARIAN199772,Aronov_1991}. Characteristic determinant approach may be considered as a ground up approach, the key idea is to start from a single cell, and gradually build up to  a many cells system by adding up one cell at a time. Using the recursion relations that determinant of $D$-matrix,    $ D=1-\hat{G}_0 \hat{V} $, satisfies, where $\hat{G}_0 = \frac{1}{E-\hat{H}_0}$ is free propagator of particle,  when the cells are periodic arranged,   scattering dynamics for a many cells system is given by   factorized two components: (1) scattering amplitudes of a single cell that is the result of short-range interaction; (2) the geometric factors that is associated with the periodic structure of  entire system, which also describe the collective modes  of finite size  crystal.

At last we remark that the compact forms of transmission amplitude and crystal-momentum $Q$ in   Eq.(\ref{tandrexpress}) and Eq.(\ref{cosQLRet0})  in terms of real and imaginary parts of $\frac{e^{- i k L} }{t_0(k)} $ are only valid and defined for the real $k$'s. For  complex $k$, e.g. in the case of looking for pole position in complex-$k$ plane,  the correct values are given by analytic continuation of  explicit expressions by inserting analytic form of $t_0(k)$  in Eq.(\ref{t0r0LReqs}) into Eq.(\ref{tandrexpress}) and Eq.(\ref{cosQLRet0}) assuming $k$ is real.

\section{Averaging tunneling time per unit cell at large \texorpdfstring{$N$}{N} limit}\label{avgtime}
As $N$ is increased, the band structure  starts showing up, and the tunneling time  starts oscillating drastically due to $\cos (Q (2N+1)L)$ and $\sin (Q (2N+1)L) $ factors. To evaluate the asymptotic behavior of tunneling time,  it is more convenient to define the tunneling time per   unit cell, 
\begin{equation}
\widehat{\tau}(k) = \frac{\tau_E}{(2N+1) L}. 
\end{equation}
For large $N$,  the second term in Eq.(\ref{integGtr}) is suppressed,  and the first term is a fast oscillating term. The result for an infinite periodic system should be approached by adding a small imaginary part to $Q$: $Q \rightarrow Q+ i \epsilon$, where $\epsilon  \gg  \frac{1}{(2 N+1) L}$. Hence the dominant term in  $t(k)$  is given by $$\sec (Q (2N+1)L) \propto e^{i Q (2N+1)L},$$ 
and 
\begin{equation}
\widehat{\tau} (k)  \stackrel{N \rightarrow \infty}{\rightarrow} i \frac{d Q}{d E}. \label{avgtauNlimit}
\end{equation}
For the  $Q$ values defined on real axis,  this conclusion may be justified by considering the averaged    tunneling time per  unit cell
\begin{equation}
\langle \widehat{\tau}(k) \rangle = \frac{1}{2 \epsilon }\int_{k-\epsilon}^{k+\epsilon} \widehat{\tau} (p)  dp  \stackrel{N \rightarrow \infty}{\rightarrow} i \frac{d Q}{d E}. \label{avgtauQ}
\end{equation}
Hence the fast oscillation is smoothed out.  Although Eq.(\ref{avgtauQ}) is not rigorously proved in this work, it can be checked rather straightforwardly in numerics, and the rigorous mathematical proof may be accomplished by using stationary phase approximation. The physical meaning of averaged    tunneling time per  unit cell may be understood as: due to limited resolution,
the experimental device usually   is only able to measure averaged result for fast oscillating objects. Fig.~\ref{tauplot1} and Fig.~\ref{tauplot2} show the typical examples of    plots of $\widehat{\tau} $ for a small size system  compared with $  i \frac{d Q}{d E}$. As we can see    in Fig.~\ref{tauplot1} and Fig.~\ref{tauplot2}   even for   a small size system  with only just a few cells, the band structures of a totally periodic system  already start building up and clearly visible. For the finite size $\mathcal{P}\mathcal{T}$-symmetric barriers, $\hat{\tau}_{1,2}$ oscillate around   asymptotic result of $ \langle \hat{\tau}_{1,2} \rangle$ at large $N$ limit  (Note that we use dimensionless quantities in all figures).

The  effect of exceptional points (EPs) that separate the broken and unbroken $\mathcal{P}\mathcal{T}$-symmetric phases in a totally periodic infinite system   \cite{10.1038/nphys4323,doi:10.1142/q0178,doi:10.1126/science.aar7709,doi.org/10.1038/s41563-019-0304-9} is also   visible in a small size system,   see the dip near $k\sim 4.1$ in Fig.~\ref{tau1plot2}. Near exceptional points, $\frac{d Q}{d E}  \sim 0$,     $ \langle \hat{\tau}_{1,2} \rangle$  approach zero,   $\mathcal{P}\mathcal{T}$-symmetric barriers become almost transparent. This phenomenon is usually referred as unidirectionally invisibility, see e.g. Ref.~\cite{PhysRevLett.106.213901,PhysRevLett.128.203904}.  The negative $\hat{\tau}_1$ near EP in Fig.~\ref{tau1plot2} is distinctive from positive tunneling time in a real potential scattering theory.  In the $\mathcal{P}\mathcal{T}$-symmetric systems, $\hat{\tau}_1$ may turn negative due to the fact that  generalized density of states of a $\mathcal{P}\mathcal{T}$-symmetric  system is not positively definite. The value of $\tau_1$ turning  negative has great deals with the spectral singularities of $\mathcal{P}\mathcal{T}$-symmetric. When the poles of $\tau_E$ move across the real axis in complex $E$-plane, they yield drastic and even divergent enhancement near spectral singularities. If  the  spectral singularities are located within the band, the crossing of poles on real axis may cause the sign flip of $\tau_1$. Following the motion of pole across real axis from unphysical sheet into physical sheet, the peak of enhancement thus moves toward positive infinity that is connected with negative infinity, and then flip the sign and continue moving away from negative infinity.  The detailed discussion about the effect of  spectral singularities is given in Sec.~\ref{sec:spectralsingularity}.

We also remark that the sign of  $Q$    in Eq.(\ref{cosQLRet0}) is not well defined and physically ambiguous. The determination of sign of $Q$ must be referred to the sign of $\hat{\tau}_E $. The   ambiguity of multivalued functions in physics in fact is  quite common. For example,  the ambiguity of    non-analyticity of scattering amplitude   in some singularities related cases has been a well-known fact in analytic $S$-matrix approach in nuclear/particle physics, see e.g. \cite{CUTKOSKY1969281,eden2002analytic}.         The nature of the non-analyticity sometime is not fully determined,  and some extra constraint must be imposed on the theory to remove the ambiguity,  for instance, using  the perturbation-theory $i \epsilon$-prescription as the reference, etc. Similarly in our case, the ambiguity of sign of $Q$ is removed by using $\hat{\tau}_E $ as the reference.

\section{Spectral singularities}\label{sec:spectralsingularity}
  With analytic expressions of transmission and reflections amplitudes given in Eq.(\ref{tandrexpress}) and Eq.(\ref{cosQLRet0}),  it can be easily checked that the traversal time  $\tau_E$ for the $\mathcal{P} \mathcal{T}$-symmetric model adopted in this work   is a well-defined analytic function in complex $E$-plane.  Two types of singularities are present:  (1) a branch cut siting along the positive real axis in complex $E$-plane   that separate physical sheet (the first Riemann sheet) and unphysical sheet  (the second Riemann sheet);    (2) poles of transmission and reflection amplitudes. These poles are called   spectral singularities of a non-Hermitian Hamiltonian  when they show up on real axis \cite{PhysRevLett.102.220402,Ahmed_2009,PhysRevB.80.165125},   which  yields   divergences of reflection and transmission coefficients of scattered states.  The spectral singularities   are interpreted as resonance states with vanishing spectral width in Ref.~\cite{PhysRevLett.102.220402}.

The motion of poles  in complex $E$-plane  has some profound impact on the tunneling  of particle through  $\mathcal{P} \mathcal{T}$-symmetric barriers. In  what follows, we will first discuss the distribution and the properties of pole singularities in Sec.~\ref{locationpoles}. How the value of $\tau_1$ is affected by the motion of poles is presented in Sec.~\ref{tau1movingpole}.  The impact of moving poles on  some other properties of tunneling time is discussed in Sec.~\ref{dispersionintegral} and Sec.~\ref{spectrallargeNlimit}.

\subsection{Distribution and motion of poles in complex plane}\label{locationpoles}

The location of these poles are model parameters dependent and can be found by solving $1/t(k)=0$. Based on Eq.(\ref{tandrexpress}),  there are two types of solutions:

(i) type I singularities are given by solutions of  $ \frac{1}{t_0 (k)} =0$.  Hence $\cos (QL) =0$ and $\frac{1}{t(k)}=0$ are both automatically satisfied, and
\begin{equation}
Q = \frac{\pi}{L}(n + \frac{1}{2})   , \ \ \ \ n\in \mathbb{Z}.
\end{equation}
The type I singularities are originated from a single cell ($N=0$),  and then carried on and shared by  the entire many cells system. The type I solutions hence are independent of number of cells and the size of system.

 \begin{figure}
\begin{center}
\includegraphics[width=0.99\textwidth]{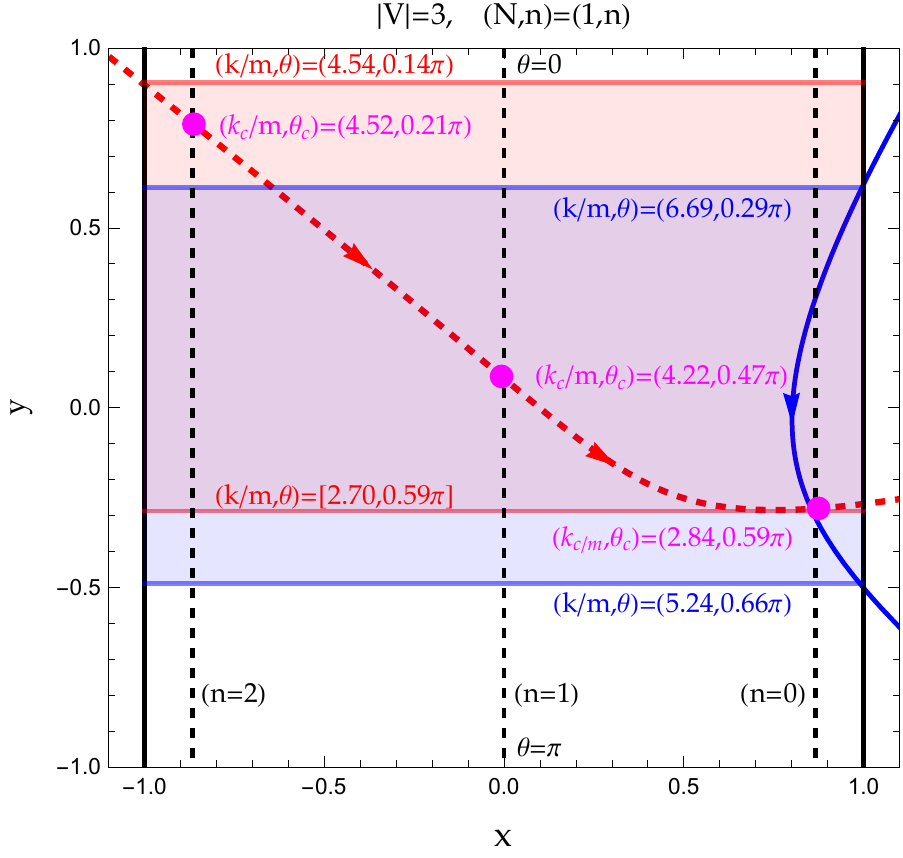}
\caption{   Spectral singularities condition plot: the parametric plot of dashed red (dashed grey) and solid blue (solid grey)  curves are generated with $(x,y)$  coordinates    given by  the left-hand side of  Eq.(\ref{spectralsinglocat})     as  a function of $k$.   The dashed black vertical line is generated with coordinates of $( \cos \left (  \frac{\pi  (n+ \frac{1}{2   }   )}{2N+1}  \right ) ,  \cos \theta)$  with $N=1$, $n=0,1,2$  and $\theta \in [0,\pi]$. The arrows  indicate increasing $\theta$ and  decreasing $  k$ directions. The value of  $(k,\theta)$   of spectral singularities for fixed $|V|$ is given by intersection  of dashed black vertical line and dashed red (dashed grey) and solid blue (solid grey) curves, and marked as purple dots(grey dots).   The $\theta$ values of  spectral singularities are indicated by $\theta_c$'s.  The shadowed bands represent allowed band of solutions as $N\rightarrow \infty$.
The dimensionless parameters are taken as:   $ |V| =3 $,  $m L=1$, $ma=0.2$ and $N = 1$. }\label{specplot}
\end{center}
\end{figure}

 \begin{figure}
\begin{center}
\includegraphics[width=0.91\textwidth]{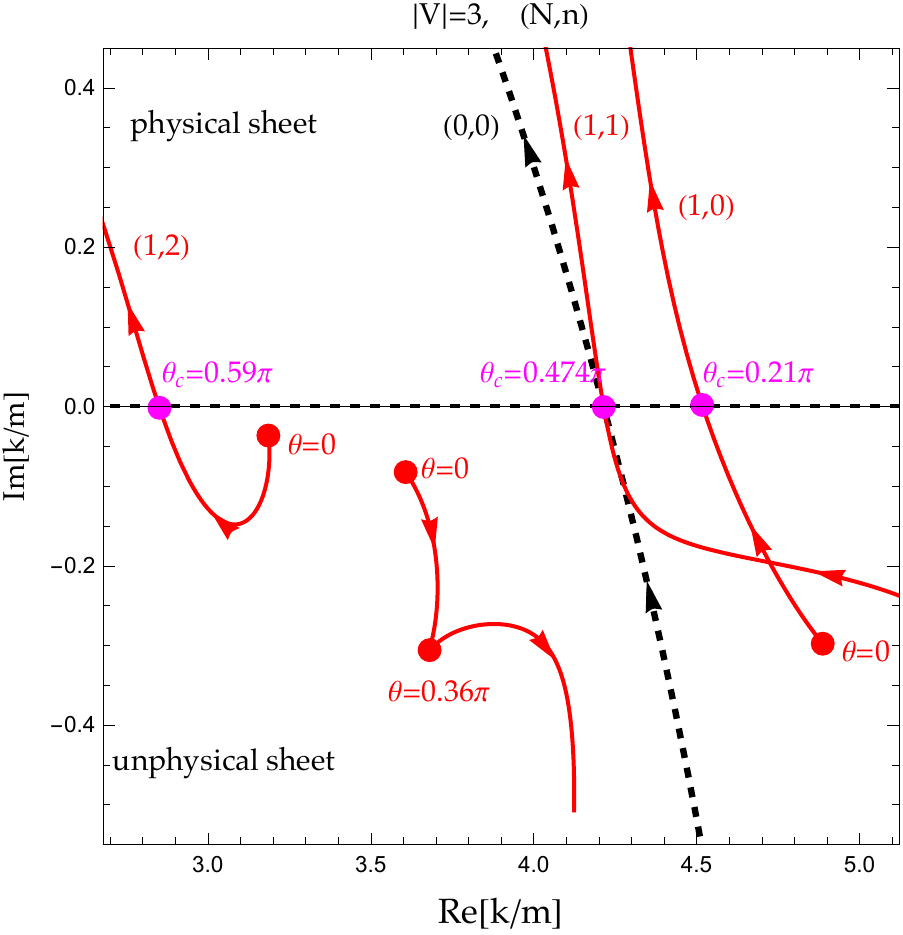}
\caption{   The motion of poles in complex $k/m$-plane as a function of increasing $\theta$ for the solutions in red band in Fig.~\ref{specplot}: $k/m \in [2.7, 4.54]$, $(N,n)=(0,0)$ (dashed black) and $(N=1,n=0,1,2)$ (solid red/solid grey).   The arrows  indicate increasing   $ \theta$ directions. The $\theta$ values of  spectral singularities are indicated by $\theta_c$'s.
The dimensionless parameters are taken as:   $ |V| =3 $,  $m L=1$, $m a=0.2$ and $N = 0,1$.}\label{motionpolesplot}
\end{center}
\end{figure}

(ii)  type II  singularities  are system size dependent and  given by  two conditions, 
\begin{equation}
\cos( Q (2 N+1) L) = 0,   \ \ \ \  Im \left [ \frac{e^{- i k L}}{t_0 (k)} \right ]=0.
\end{equation}
Using Eq.(\ref{t0r0LReqs}) and Eq.(\ref{cosQLRet0}),  two conditions can be rearranged to  the form
\begin{equation}
\left  (x (k) , y (k) \right ) = \left ( \cos \left (  \frac{\pi  (n+ \frac{1}{2   }   )}{2N+1}  \right ) , \cos \theta \right ), \label{spectralsinglocat}
\end{equation}
where
\begin{align}
& x(k) =  \frac{1- 2 \left [ \frac{m |V|}{k} \sin (k 2 a)  \right  ]^2  }{ \cos (k L)}      , \nonumber \\
& y(k) =        \frac{  1- 2 \left [ \frac{m |V|}{k} \sin (k 2 a)  \right  ]^2    }{2  \frac{  m |V|}{k}  \cot  (k L)   } -  \frac{1}{2} \frac{m |V|}{k}  \sin (k 4 a)    .
\end{align}
The  $2N+1$   independent integer $n$'s  are labeled as
$$n=0,1,\cdots, 2N,$$ so $Q=\frac{\pi  (n+ \frac{1}{2   }   )}{(2N+1)L} $ sit in the first Brillouin zone.   In fact type I singularity solutions on real axis coincide with    solutions of  Eq.(\ref{spectralsinglocat})   with    $n=N$, hence $Q= \frac{\pi}{2 L}$ and   $\cos (QL) =0$  as well for type II   solutions with $n=N$. However, type I and type II solutions of $n=N$     diverge in complex plane, see e.g. Fig.~\ref{motionpolesplot}.

  The solutions of   spectral  singularities (poles on real axis) can be visualized graphically by plotting parametric curves using both sides of Eq.(\ref{spectralsinglocat}) as  $(x, y)$ coordinates for a fixed $|V|$.  The intersections of curves indicate the location of  spectral  singularities,  see e.g. Fig.~\ref{specplot}. The distribution of spectral singularity solutions are split up and poles are clustered into bands with gaps in between,  see Fig.~\ref{specplot}, where  lowest two bands, $k/m \in [2.7,  4.52]$ and $k/m \in [5.24, 6.69]$,   are marked as shadowed areas with red and blue   colours (different grey scale colours) respectively. As $N$ is increased, the number of solutions in each band grow linearly. 
  The motion of poles in complex $k$-plane is illustrated in Fig.~\ref{motionpolesplot}. For $\theta \sim 0$, all the poles are located in unphysical sheet (the second Riemann sheet). As $\theta$ is increased, some poles start moving across real axis into physical sheet (the first Riemann sheet). The critical value of $\theta_c$ for spectral singularities is individually dependent, see  e.g. Fig.~\ref{motionpolesplot}. The density of solutions over small  $\theta$ interval is controlled by the inverse slope of dashed red (dashed grey) and solid blue (solid grey) curves in Fig.~\ref{specplot}, the flatter the curves are, the more number of spectral  singularities are going to cross the real axis over a small range of $\theta$ increment.

 In addition,  it is also easy to show, using  Eq.(\ref{spectralsinglocat}), that for the large   $|V|$ the poles are located around the zeros of ${\sin (2ka)}$.
At the limiting case $|V|\rightarrow \infty$,   poles start approaching   $k=\frac{\pi n}{2a}$. The physical meaning of such coincidences can be understood by relating it to the formation of resonant states in a single cell,  where the distance between two delta potentials $2a$ now plays a dominant role, rather than the length of the cell $L$.
As a consequence,  the electron spends most of its time  moving back and forth   before leaving the cell. With decreasing $|V|$,  the poles start moving toward large $k$, and number of poles at low energy region   decrease and approaches zero.
Note that in the limit $kL\ll 1$ and in the first-order Taylor expansion, the positions of the poles do not depend on $k$ and the following approximate expression 
$m |V_{cr}|\approx \frac{1}{2a}\sin \frac{\pi}{4}\frac{2n+1}{2N+1}
$
can be used to calculate the critical value $|V_{cr}|$ for given parameters: $N$, $n$ and $a$.

  \begin{figure*}
 \centering
 \begin{subfigure}[b]{0.49\textwidth}
\includegraphics[width=0.99\textwidth]{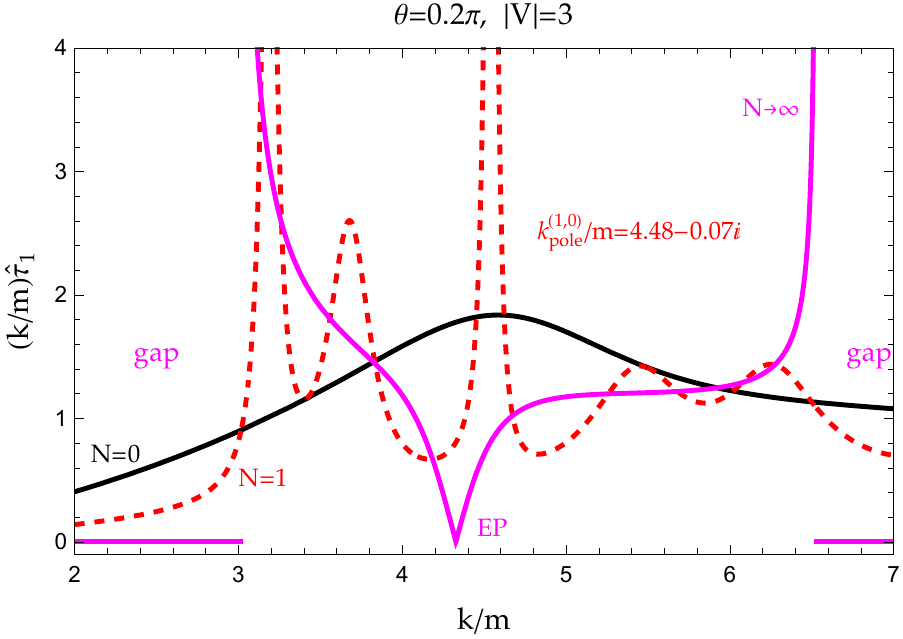}
\caption{   }\label{tau1V3plot1}
\end{subfigure} 
\begin{subfigure}[b]{0.49\textwidth}
\includegraphics[width=0.99\textwidth]{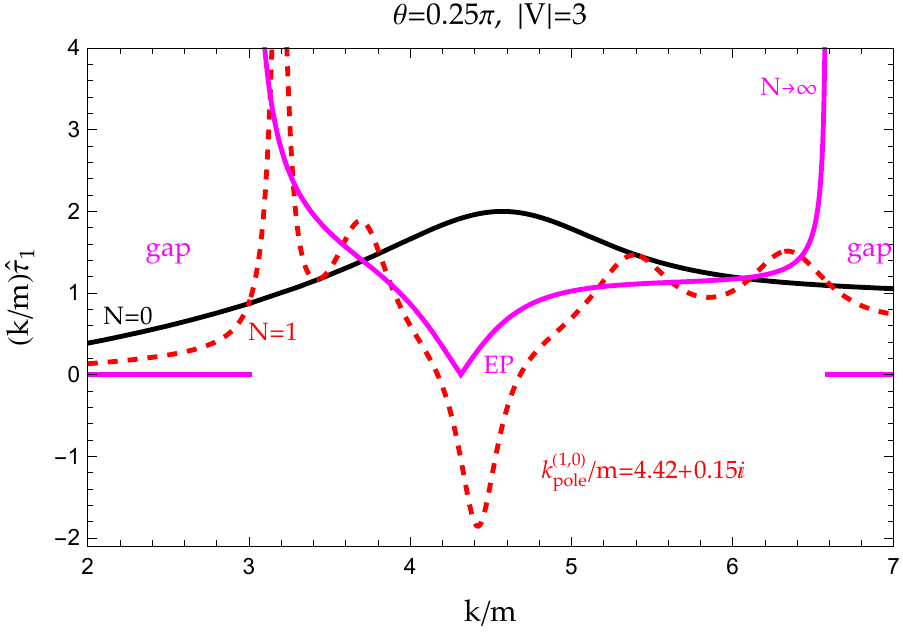}
\caption{     }\label{tau1V3plot2}
\end{subfigure}
\begin{subfigure}[b]{0.49\textwidth}
\includegraphics[width=0.99\textwidth]{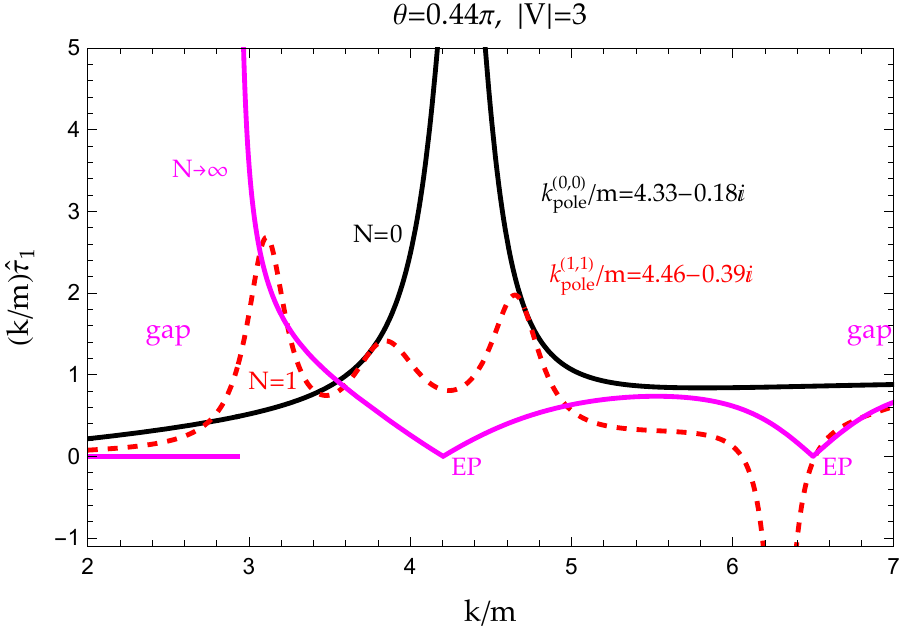}
\caption{  }\label{tau1V3plot3}
\end{subfigure}
\begin{subfigure}[b]{0.49\textwidth}
\includegraphics[width=0.99\textwidth]{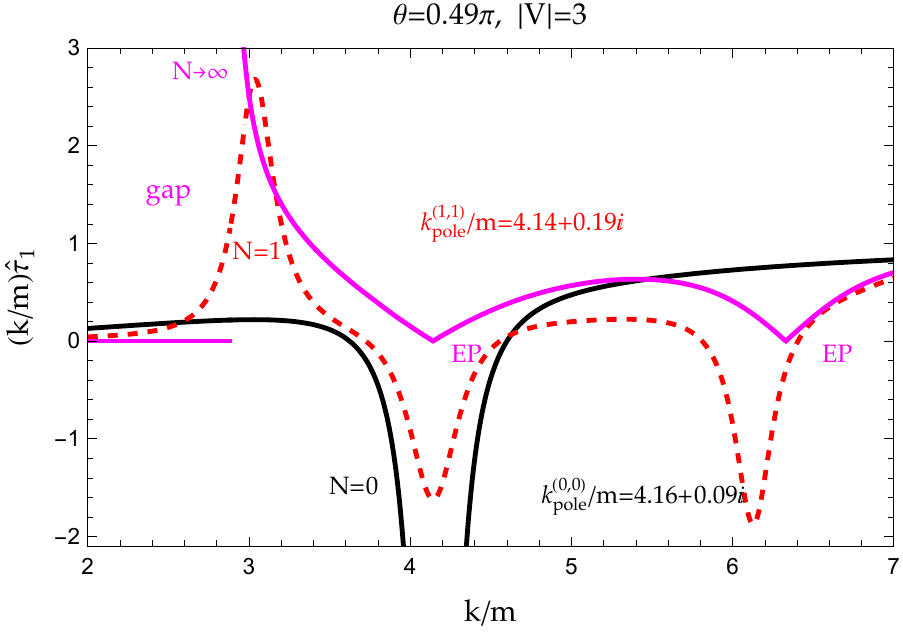}
\caption{   }\label{tau1V3plot4}
\end{subfigure}
\begin{subfigure}[b]{0.49\textwidth}
\includegraphics[width=0.99\textwidth]{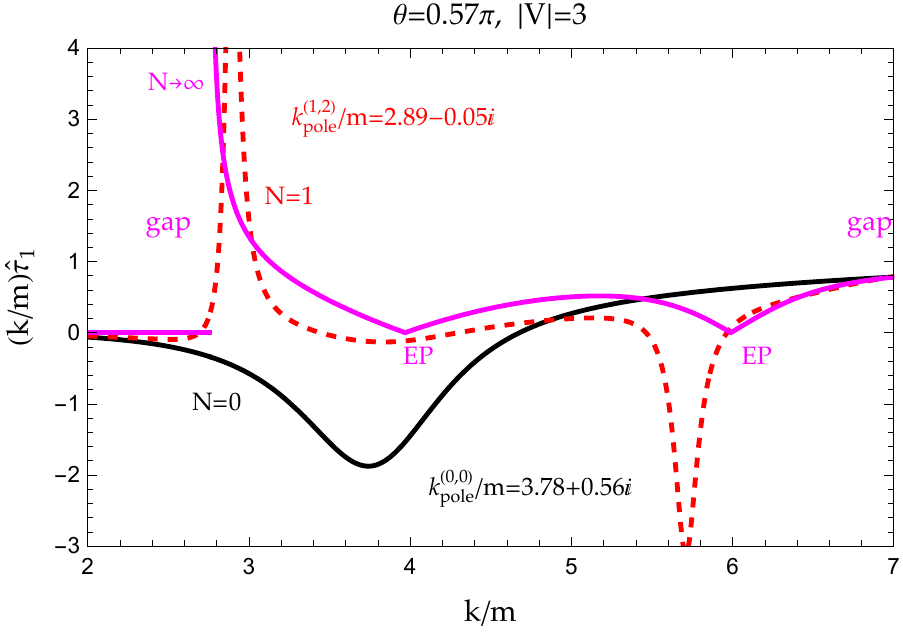}
\caption{ }\label{tau1V3plot5}
\end{subfigure}
\begin{subfigure}[b]{0.49\textwidth}
\includegraphics[width=0.99\textwidth]{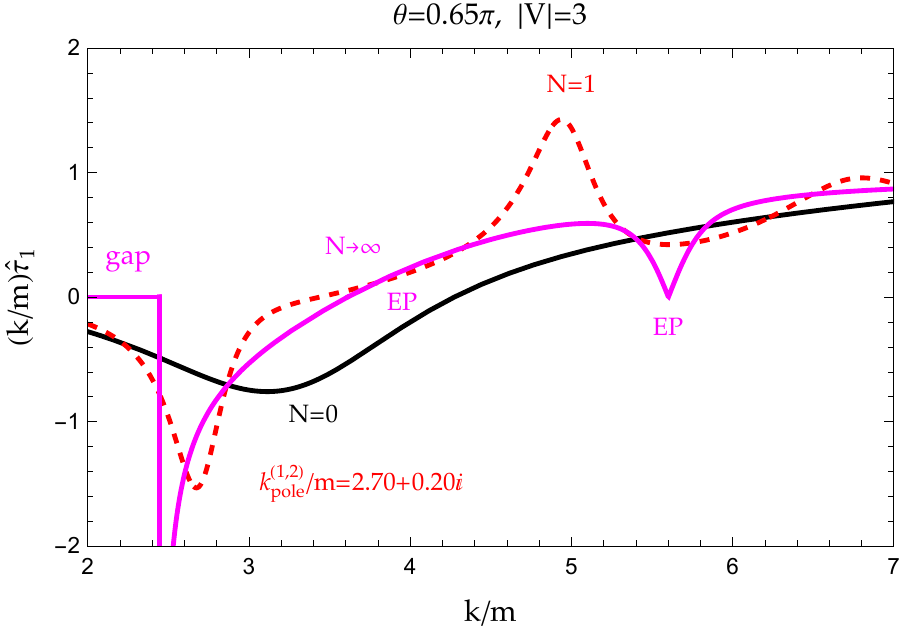}
\caption{  }\label{tau1V3plot6}
\end{subfigure}
\caption{  Plot of  $\frac{k}{m} \widehat{\tau}_1$ with $N=0$ (solid black), $N=1$ (dashed red/dashed grey)  together with   $   \frac{k}{m}   \frac{d Re[Q ] }{d E}  $    (solid purple/light grey)  vs. $k/m$ for various $\theta$ values: (a)  $\theta =0.2 \pi$; (b) $\theta =0.25  \pi$; (c) $\theta =0.44 \pi$; (d) $\theta =0.49 \pi$; (e) $\theta =0.57 \pi$; (f) $\theta =0.65 \pi$. Some of pole positions near real axis are listed and marked in red (grey) ($N=1$) and black ($N=0$). The dimensionless parameters are taken as:    $|V| =3 $, $m L=1$ and $m a=0.2$. }\label{tau1V3plots} 
\end{figure*}

  \begin{figure*}
 \centering
 \begin{subfigure}[b]{0.49\textwidth}
\includegraphics[width=0.99\textwidth]{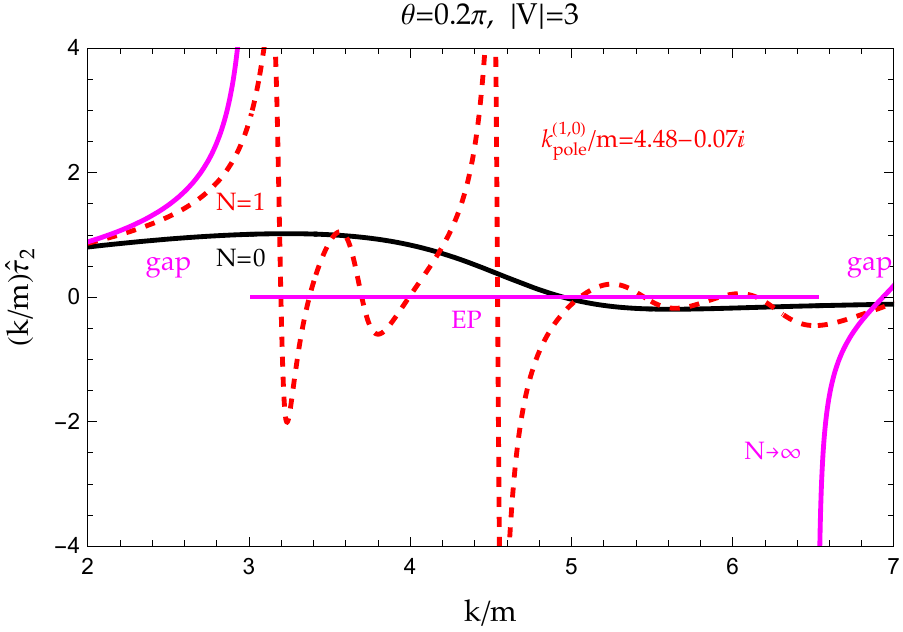}
\caption{   }\label{tau2V3plot1}
\end{subfigure} 
\begin{subfigure}[b]{0.49\textwidth}
\includegraphics[width=0.99\textwidth]{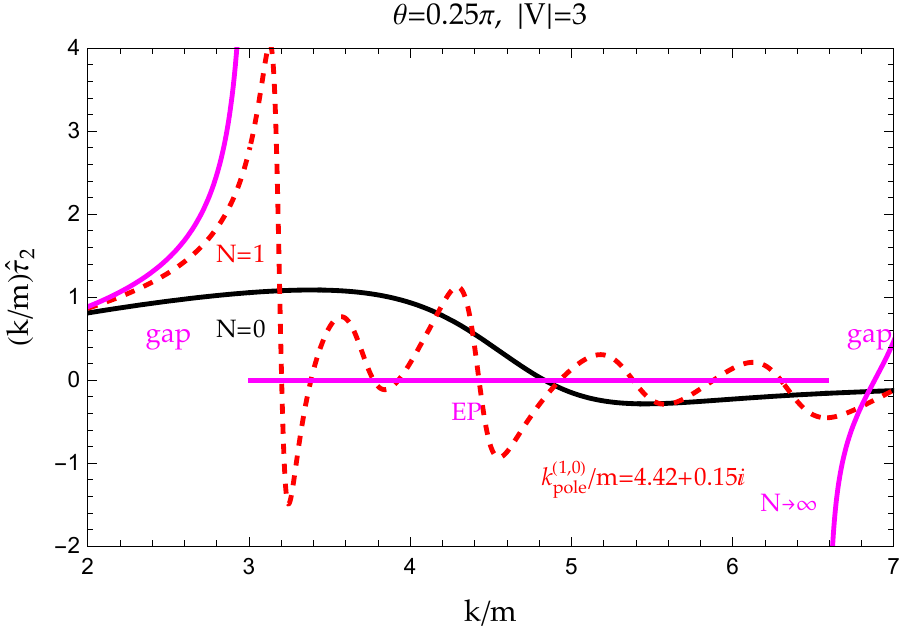}
\caption{     }\label{tau2V3plot2}
\end{subfigure}
\begin{subfigure}[b]{0.49\textwidth}
\includegraphics[width=0.99\textwidth]{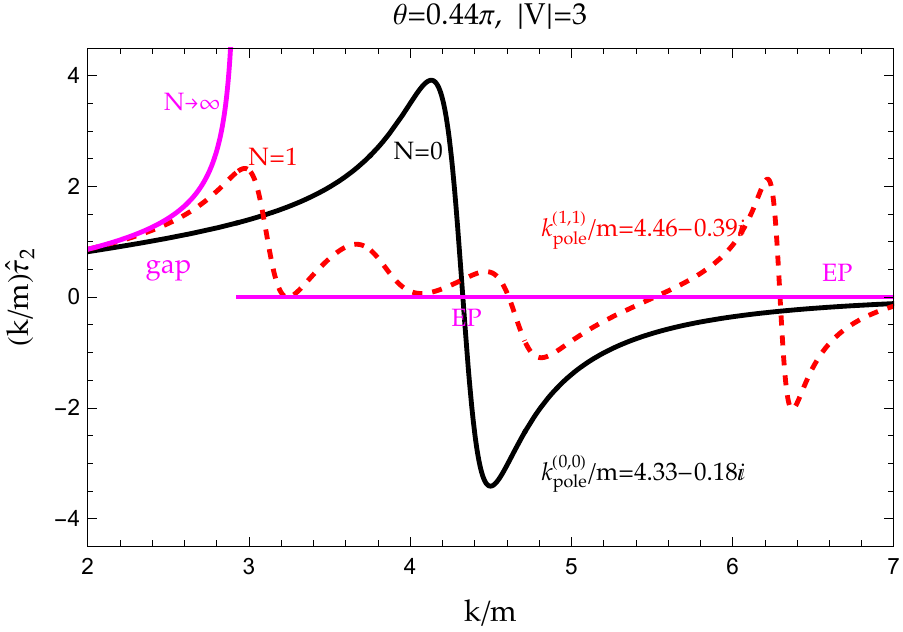}
\caption{  }\label{tau2V3plot3}
\end{subfigure}
\begin{subfigure}[b]{0.49\textwidth}
\includegraphics[width=0.99\textwidth]{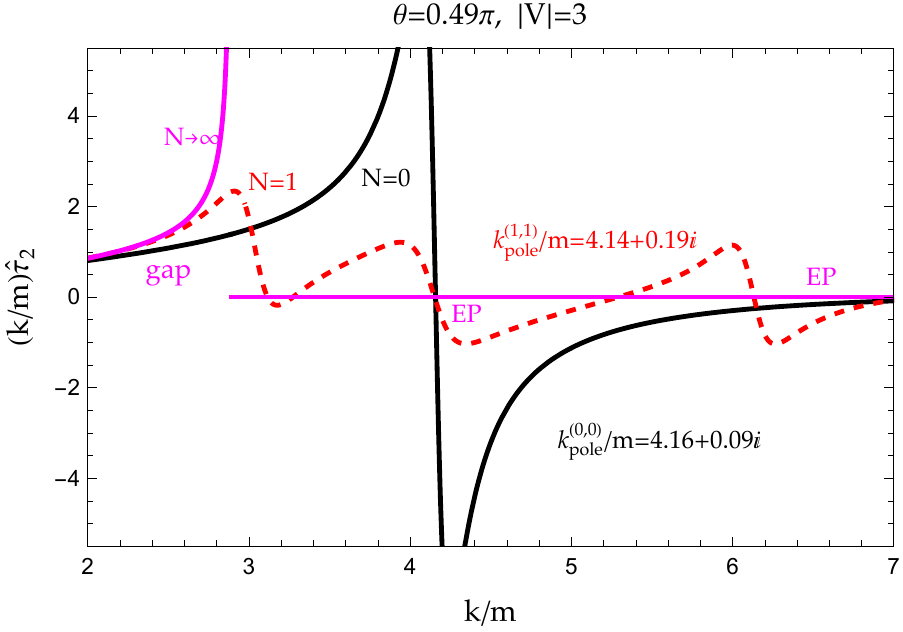}
\caption{   }\label{tau2V3plot4}
\end{subfigure}
\begin{subfigure}[b]{0.49\textwidth}
\includegraphics[width=0.99\textwidth]{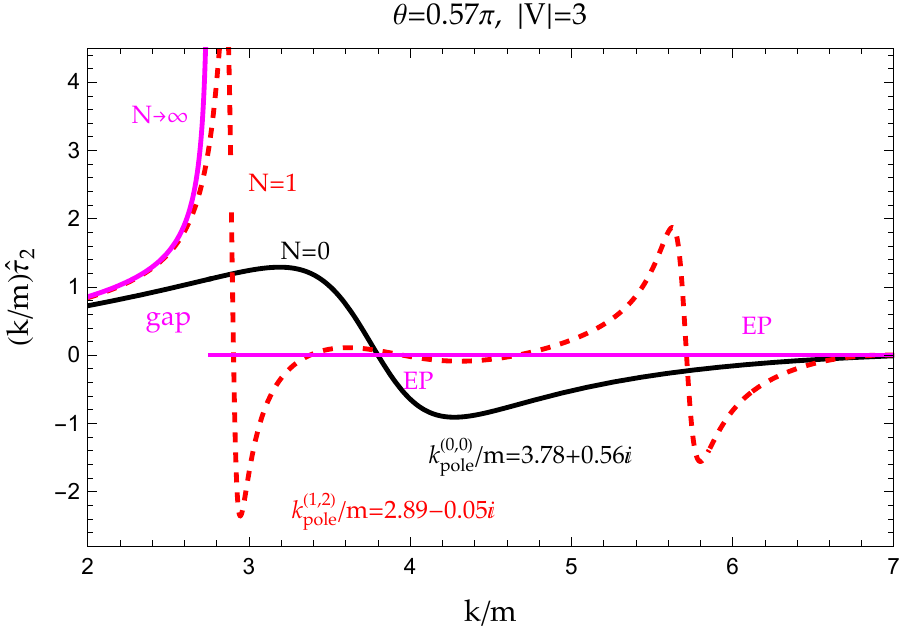}
\caption{  }\label{tau2V3plot5}
\end{subfigure}
\begin{subfigure}[b]{0.49\textwidth}
\includegraphics[width=0.99\textwidth]{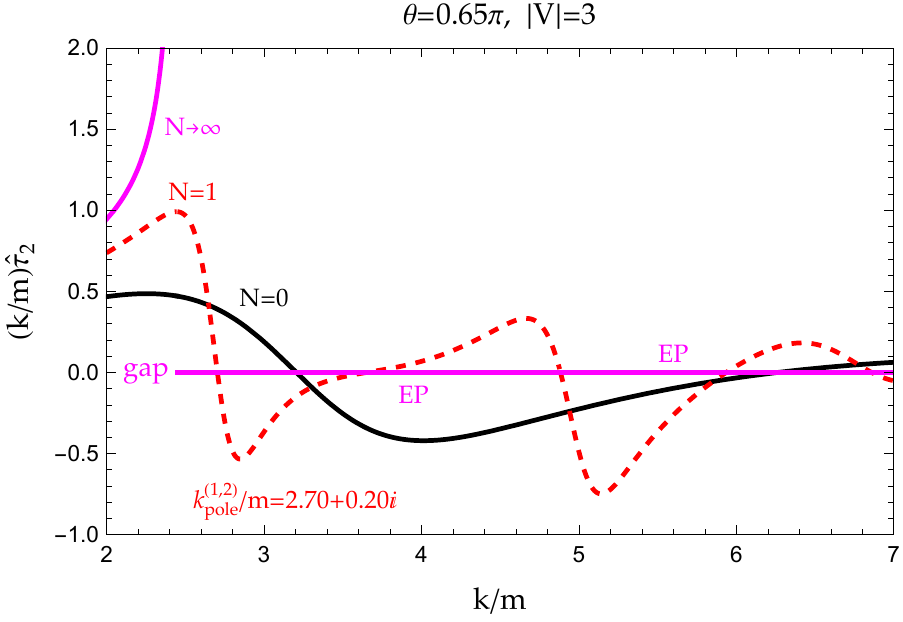}
\caption{  }\label{tau2V3plot6}
\end{subfigure}
\caption{  Plot of  $\frac{k}{m} \widehat{\tau}_2$ with $N=0$ (solid black), $N=1$ (dashed red/dashed grey) together with  $   \frac{k}{m}   \frac{d Im[Q ] }{d E}  $ (solid purple/light grey)  vs. $k/m$, for various $\theta$ values: (a)  $\theta =0.2 \pi$; (b) $\theta =0.25  \pi$; (c)  $\theta =0.44 \pi$; (d) $\theta =0.49 \pi$; (e) $\theta =0.57 \pi$; (f) $\theta =0.65 \pi$. Some of pole positions near real axis are listed and marked in red (grey) ($N=1$) and black ($N=0$). The dimensionless parameters are taken as:     $|V| =3 $, $m L=1$ and $m a=0.2$. }\label{tau2V3plots} 
\end{figure*}

\subsection{Negative \texorpdfstring{$\tau_1$}{tau1} and its relation to moving poles}\label{tau1movingpole}

In the real potential scattering, $\tau_1$ always remain positive because of the positivity of density of states of systems. However, in  $\mathcal{P}\mathcal{T}$-symmetric systems, $\tau_1$ now is related to the generalized density of  states, which could be either positive or negative and so is $\tau_1$.   For $\theta \sim 0$,    $\mathcal{P}\mathcal{T}$-symmetric systems behave just like normal real potential system, $\tau_1$ remains positive. As $\theta$ is increased,   the value of $\tau_1$ may turn negative at certain   energy ranges.   $\tau_1$  turning negative  is closely related to the motion of poles across the real axis moving  from unphysical sheet (the second Riemann sheet) into physical sheet (the first Riemann sheet).

For   a single cell   $(N=0)$, no band structure can be observed yet. The situation is relatively simple, every time when the pole  crosses real axis and moves into physical sheet, the value of  $\tau_1$  turns negative near the crossing points. Example   is illustrated  in Fig.~\ref{tau1V3plots}.  For a single cell,  only a single spectral singularity can be found near $k/m=4.22$ at $\theta_c=0.474\pi$, see Fig.~\ref{motionpolesplot}. As we can see in  Figs.~\ref{tau1V3plot1}-\ref{tau1V3plot3}, for $\theta < \theta_c = 0.474\pi$, the pole is still located in unphysical sheet, and  $\tau_1$'s  values (solid black) are positive.  As $\theta $ approaches   $  \theta_c = 0.474\pi$, the pole on unphysical sheet moves  towards real axis,  and the peak of enhancement in $\tau_1$ that is generated by the pole moves up to positive infinity.  The  spectral singularity at $\theta=\theta_c$ is  a critical point where the peak of   positive infinity   meets  negative infinity.  As   $\theta $  is increased   and passes over $ \theta_c  $, see  Figs.~\ref{tau1V3plot4}-\ref{tau1V3plot6}, the pole   now already moves across real axis into physical sheet, and the peak in $\tau_1$  moves across the boundary between positive and negative infinity and turns negative.     As    $\theta $ is continuously increased, the pole moves away from real axis on physical sheet,   which ultimately yields a negative bump near pole location in   $\tau_1$. The steepness of bump is determined by how close the pole is to the real axis. This can be easily demonstrated   with the motion of  a single pole. Near the pole, the transmission amplitude is approximated by
\begin{equation}
t(k) \propto \frac{1}{k-k_{pole}  } = \frac{k-k_{re} - i  \epsilon}{ (k-k_{re})^2+ \epsilon^2}, 
\end{equation}
where  $k_{pole} = k_{re} + i \epsilon$,  being $k_{re}$ and $\epsilon$ the real and imaginary parts of pole position. The location of  pole   in physical sheet  or  unphysical sheet is determined by sign of $\epsilon$:  unphysical sheet if $\epsilon <0$ and physical sheet if $\epsilon >0$. The $\tau_1$ near the pole is thus dominated by
\begin{equation}
\tau_1 \sim \frac{m}{k} \frac{   \epsilon}{ (k-k_{re})^2+ \epsilon^2},
\end{equation}
hence as pole moves across real axis into physical sheet, $\epsilon$ changes  its sign.

In the case of $N=1$, the size of system is still small, but the band structure already  starts showing up. Three spectral singularity solutions  can be found within the first cluster of solutions in $k/m \in [2.7,  4.52]$, see    Fig.~\ref{specplot} and Fig.~\ref{motionpolesplot}. The locations are $k/m = (4.52, 4.22, 2.84)$ at $\theta_c = (0.21 \pi, 0.47 \pi, 0.59 \pi)$ respectively.  When the first pole  crosses the real axis at $(k/m,\theta_c) =(4.52, 0.21\pi)$ and moves into physical sheet, it generates a negative bump in $\tau_1$, see  Fig.~\ref{tau1V3plot2}. As $\theta $ is continuously increased  near $\theta_c =0.47\pi$, second pole is  getting close to  real axis at $k/m=4.22$ and  starts competing with the nearby first pole in physical sheet. It becomes dominant near  $\theta_c =0.47\pi$, and turns $\tau_1$ back into positive, see  Fig.~\ref{tau1V3plot3}. After the second pole crosses the real axis, it  flips $\tau_1$ again   and generates a negative bump at $k/m\sim 4.14$, see  Fig.~\ref{tau1V3plot4}. Similarly as $\theta$ is continuously increased up to $\theta_c= 0.59 \pi$, the last pole  moves in and becomes the dominant effect  in turning $\tau_1$  when it  crosses the real axis near  $k/m\sim2.7$, see   Fig.~\ref{tau1V3plot5} and   Fig.~\ref{tau1V3plot6}.

For a large $N$ system, the situation is even more interesting,   the band structure and EPs     start getting involved, competing with poles and playing the roles in turning $\tau_1$. The spectral singularities are clustered into bands, the first band shows up in $k/m  \in [2.7,  4.52]$ at corresponding $ \theta \in [ 0.59 \pi,  0.21\pi]$. For $\theta  \in [ 0.21 \pi, 0.5\pi ]$, spectral singularities and poles are clustered in $k/m \in [ 4, 4.5]$, which happens near the edge of merging two bands (EPs). The effect of poles located near the edge of two merging bands is hence highly suppressed by EPs, the $\tau_1$ remain positive.   For $\theta  \in [0.5 \pi, 0.59\pi ]$, the cluster of spectral singularities and poles are now moved up to lower edge of band $k/m \sim 2.8$,  they become the dominant force of determining the fate of $\tau_1$. When they move  across the real axis, the  $\tau_1$ in lower band below EP is turned completely, see purple (light grey) curves in   Fig.~\ref{tau1V3plot5} and   Fig.~\ref{tau1V3plot6}.

Now the physical interpretation of negative value of $\tau_1$ as repelling time that is physically inaccessible can be understood in terms of the motion of poles. As we can see in  Fig.~\ref{tau1V3plots}, the turning  to negative of $\tau_1$ is closely related to the motion of pole crossing real axis.  As the pole  crosses real axis from unphysical sheet into physical sheet, the value of $\tau_1$ experiences the transition process from positive value  to positive infinity that is connected with negative infinity, and ultimately to negative value. $\tau_1$  diverges at spectral singularity and generates an infinite sharp barrier for particle to pass through. As the pole moved into physical sheet, then the barrier is broadened and turned into a negative value band that repel particle. The behavior of $\tau_2$ is plotted in Fig.~\ref{tau2V3plots}.

 \begin{figure}
\begin{center}
\includegraphics[width=0.99\textwidth]{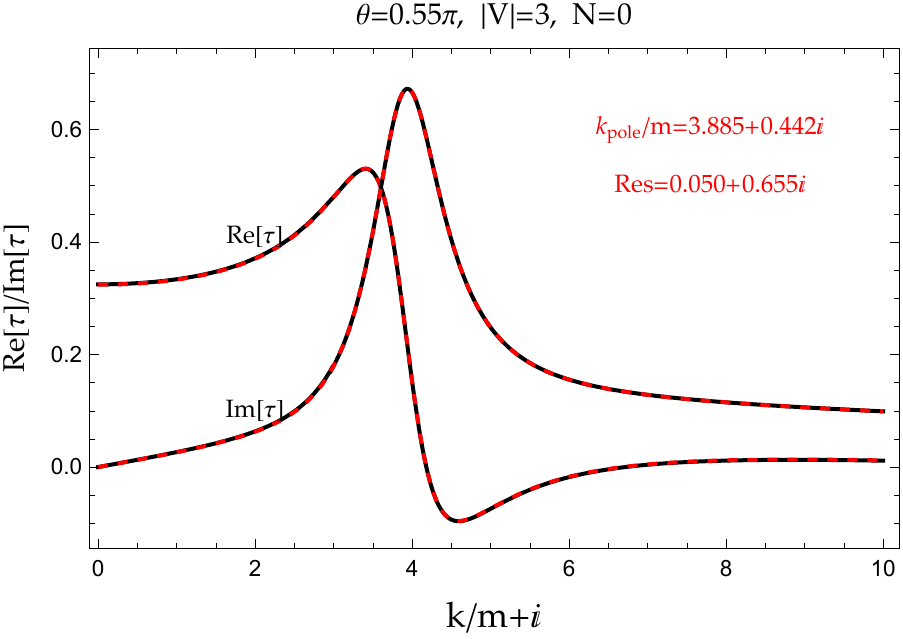}
\caption{  The plot of real and imaginary parts of $\tau_E$ by dispersion integral relation in Eq.(\ref{polecauchyinteg}) (solid black)  together with $\tau_E$  (dashed red/grey) for $N=0$  (the curves overlap). Only one pole is present at $k/m=3.885+0.442 i$.    $\tau_E$ is computed  and shown off real $E$ axis only for the purpose of fast convergence of Cauchy integral. 
The parameters are taken as: $\theta =0.55 \pi$,  $ |V| =3 $,  $m L=1$ and $m a=0.2$, where $|V|$ is dimensionless.}\label{taudispplot}
\end{center}
\end{figure}

\subsection{Dispersion integral relation of \texorpdfstring{$\tau_E$}{tauE}}\label{dispersionintegral}

 In addition to turning the value of $\tau_1$, the motion of pole singularities also has a big impact on the  dispersion integral relation of $\tau_E$. For the small  $\theta \sim 0$, all the spectral singularities are located in unphysical sheet   or equivalently  in the lower half  complex $k$-plane with $Im [k] <0$.    Hence, except the branch cut, no other singularities can be found in physical sheet, the  $\tau_E$ must satisfies Cauchy's integral theorem (also referred as dispersion integral relation in nuclear/particle physics),
  \begin{equation}
  \tau_E = \frac{1}{\pi} \int_0^\infty d \omega \frac{Disc_\omega \tau_\omega }{\omega - E}  , \label{cauchyinteg}
  \end{equation}
  where for $\mathcal{P}\mathcal{T} $-symmetric system, the discontinuity of $\tau_E$ crossing the branch cut is $$Disc_E \tau_E = Im [\tau_E] = \tau_1 (E)$$ for real values of $E$.  Specifically for the model used in this work, asymptotically $\tau_1(E)  \rightarrow 0$  as either $E \rightarrow 0$ or $E \rightarrow \infty$, hence the Cauchy's integral on the right-hand side of Eq.(\ref{cauchyinteg}) is well-behaved and converging. No subtractions or extra constant terms are needed.

As $\theta$ is continuously increased,  poles of transmission and reflections amplitudes start moving around in complex plane, some move across the branch cut on real axis and start interfere with Cauchy integration contour, see e.g. Fig.~\ref{motionpolesplot}. When poles   from unphysical sheet  move across branch cut into physical sheet,   the contour of Cauchy integral is dragged to follow the motion of poles and move with poles together. The residue contribution of the poles in physical sheet thus must be picked up due to the deformation of the contour of Cauchy integral, hence Eq.(\ref{cauchyinteg}) is modified to
    \begin{align}
  \tau_E &= \sum_i  \left [\frac{ 2 i Res_{pole-i}}{E- E_{pole-i} } -\frac{ 2 i {Res}^*_{pole-i} }{E-  E_{pole-i}^* }  \right ] \nonumber \\
  &  + \frac{1}{\pi} \int_0^\infty d \omega \frac{Disc_\omega \tau_\omega }{\omega - E}  , \label{polecauchyinteg}
  \end{align}
 where $E_{pole-i}$ stands for the position of $i$-th pole.    The residue of $i$-th pole of $Disc_E \tau_E$ function,  $Res_{pole-i}$,  is    given by
 \begin{equation}
 Res_{pole-i} =(E-E_{pole-i} )  Disc_E \tau_{E } |_{E \rightarrow E_{pole-i} }.
 \end{equation} 
 For the complex $E$,    $Disc_E \tau_E$ function must be generalized to
  \begin{equation}
   Disc_E \tau_{E }  = \frac{\tau_E - \tau_{E^*}}{2i} .
 \end{equation} 
   The poles of transmission and reflections amplitudes in $\mathcal{P}\mathcal{T}$-symmetric systems always show up in complex conjugate pairs. In terms of momenta, the conjugate pair, $E_{pole-i}$ and $E^*_{pole-i}$, are associated  with $k_{pole-i}$ and $- k_{pole-i}^*$ respectively.   The  pair of conjugate pole terms together guarantees that the pole terms all together in Eq.(\ref{polecauchyinteg})  are always real for   $E$  on real axis.
 Eq.(\ref{polecauchyinteg}) can be checked numerically rather straightforwardly for small $N$ system, see e.g. Fig.~\ref{taudispplot}.  As $N$ grows, the number of poles soon become too large to manage.

  \begin{figure*}
 \centering
 \begin{subfigure}[b]{0.49\textwidth}
\includegraphics[width=0.99\textwidth]{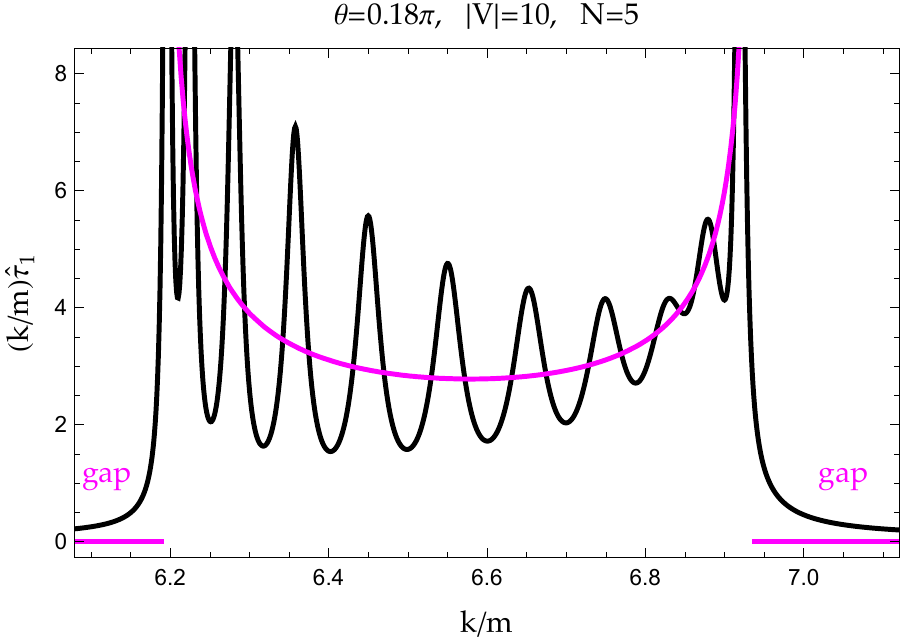}
\caption{   }\label{tau1V10plot1}
\end{subfigure} 
\begin{subfigure}[b]{0.49\textwidth}
\includegraphics[width=0.99\textwidth]{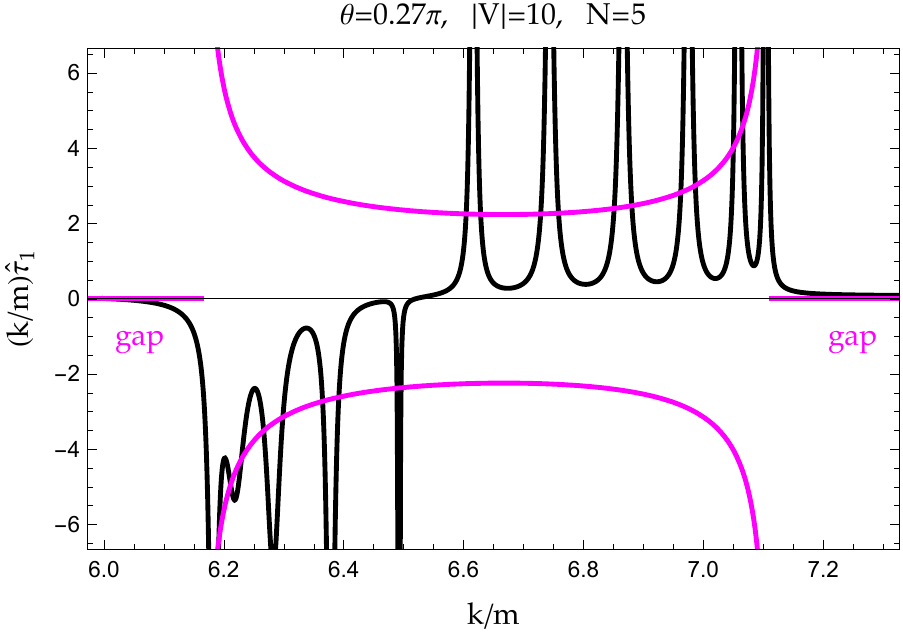}
\caption{     }\label{tau1V10plot2}
\end{subfigure}
\begin{subfigure}[b]{0.49\textwidth}
\includegraphics[width=0.99\textwidth]{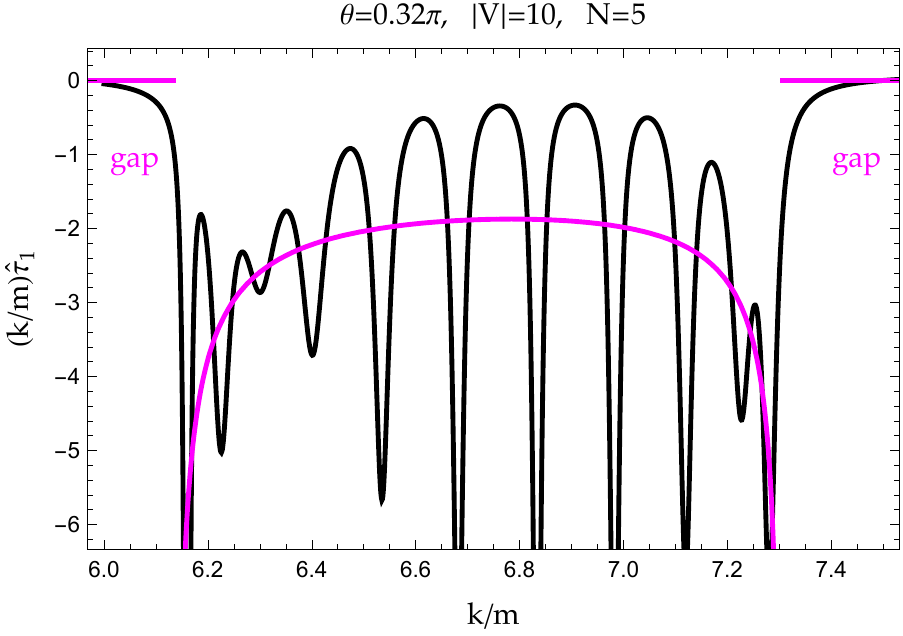}
\caption{  }\label{tau1V10plot3}
\end{subfigure}
\caption{   Plot of  $\frac{k}{m} \widehat{\tau}_1$ with $N=5$ and  $|V| =10 $ (solid black) together with   $   \frac{k}{m}   \frac{d Re[Q ] }{d E}  $  (solid purple/light grey) for various $\theta$ values: (a) $\theta =0.18 \pi$; (b) $\theta =0.27  \pi$; (c) $\theta =0.32 \pi$. The rest of parameters are taken as:       $m L=1$ and $ma=0.2$. }\label{tau1V10plots}  
\end{figure*}

\section{Discussion and summary}\label{summary}

\subsection{Large \texorpdfstring{$N$}{N} limit in presence of spectral singularities?}\label{spectrallargeNlimit}

  The averaging tunneling time in  Eq.(\ref{avgtauNlimit})  works well and  is mathematically well-defined in bands where spectral singularities are absent on real axis, the poles are all either in physical sheet or already all crossed real axis into unphysical sheet. However, in the bands where   the divergent singularities show up on the real axis and the band is still in the middle of transition between all positive and all negative band of $\tau_1$, see e.g. Fig.~\ref{tau1V10plots},  Eq.(\ref{avgtauNlimit}) break down,  and large $N$ limit becomes ambiguous and problematic.  Now we are facing the question:  is there a physically meaningful large $N$ limit in the region where divergent singularities show up?   Should we even bother to ask such a question? Or perhaps the size dependent fast oscillating and divergent behavior in   $\tau_1$ region is the nature of  $\mathcal{P}\mathcal{T}$-symmetric systems? If such a limit indeed  exists, should   the size dependent type II spectral singularities be all smoothed and washed out? Is a large size periodic system supposed to be   free of type II spectral singularities?  We still do not have a clear answer to these questions.

One very interesting observation is that as $N \rightarrow \infty$, solutions of poles are pushed either towards the real axis or pushed far away from real axis.  This can be understood that the pole position depends on factors such as $e^{ i Q (2N+1)L} $ and  $e^{  i k (2N+1)L} $, hence as   $N \rightarrow \infty$, two factors are either  fast grow or decay exponentially  for $Q$ and $k$ in complex plane. The divergence of spectral singularities may be washed out by computing value of $\hat{\tau}( k \pm i \epsilon) $ away from singularities   where $\epsilon  $ is much larger than the imaginary part of pole solutions. However, similarly to the ambiguity in sign of $Q$, for  the band that is still in the middle of transition, some parts of band have already turned negative but other parts remain positive, now we are facing the ambiguity of shifting $k$ above or below singularities.  In addition, one may also wonder how the nature  gain the access to the energy in complex plane. The idea proposed by Lloyd in Ref.~\cite{Lloyd_1969} may shed some light on this inquiry: the averaged Green's function of the  disordered system with uncorrelated disorder of Cauchy type is equal to  the Green's function of the ordered system with the energy argument shifted into complex plane.

\subsection{Summary}

In summary, the concept of tunneling time is generalized and applied to the $\mathcal{P}\mathcal{T}$-symmetric systems. The distinctive features and properties of tunneling time in $\mathcal{P}\mathcal{T}$-symmetric systems are studied and discussed by using a simple exactly solvable diatomic $\mathcal{P}\mathcal{T}$-symmetric impurities model. Unlike the positively definite $\tau_1$ in real potential scattering theory, the $\tau_1$ in $\mathcal{P}\mathcal{T}$-symmetric systems can be either positively or negatively valued. The value of $\tau_1$ turning negative is closely related to the motion of pole singularities of scattering amplitudes in complex $k$-plane. When the poles are all located in unphysical sheet (the second Riemann sheet), $\tau_1$ remains positive. As poles moves close to and ultimately  cross real axis into physical sheet (the first Riemann sheet), the poles generate a   enhancement in $\tau_1$ near the location of poles. The peak of enhancement moves into positive infinity and then moves back in from negative infinity with the sign flipped. For the large size systems, the situation is even more intriguing. The band structure of system is clearly visible for even small size systems. The number of poles grow drastically with size, and the distribution of poles split into bands. When the poles show up inside an allowed band of system and all move across the real axis, they tend to flip the sign of entire band. In some bands where  two bands start merging together at exceptional point, the exceptional points tend to force $\tau_1$ approaching zero and starts to competing with poles, so the $\mathcal{P}\mathcal{T}$-symmetric  systems become almost transparent near EPs. The fate of $\tau_1$ near EPs now is  the result of two competing forces: the poles and EPs.

 The negative value  of $\tau_1$ is distinctive feature of  $\mathcal{P}\mathcal{T}$-symmetric systems as the consequence of norm violation, hence it may   be used to quantify and calibrate the degree of norm violation in $\mathcal{P}\mathcal{T}$-symmetric systems even with balanced gain and loss.  In addition, since  negative value portion  of $\tau_1$  is physically inaccessible and  behave just like a forbidden gap in conventional periodic real potential systems. Therefore,  by manipulating balanced gain and loss in a $\mathcal{P}\mathcal{T}$-symmetric  system, we may be able to manufacture new type of band structure electronic or optical devices  even for a small finite size non-periodic system.

The impact of spectral singularities on  dispersion integral relation of $\tau_E$ and large $N$ limit is also discussed. As poles move across real axis into physical sheet, the contour of Cauchy integral must be deformed and follow the motion of poles, hence the residue terms must be picked up in dispersion integral relation of $\tau_E$. In the band of absent spectral singularities, the large $N$ limit is well defined and can be achieved by either averaging fast oscillating behavior of $\tau_E$ or using $i \epsilon$-prescription  by shifting $k$  off real axis into complex plane. However, in the bands that are plagued by divergent spectral singularities, defining a large $N$ limit becomes  problematic. The question of how to define a physically meaningful large $N$ limit in presence of spectral singularities  is still open.

\acknowledgments

P.G. and V.G.   acknowledge support from the Department of Physics and Engineering, California State University, Bakersfield, CA.  V.G. and E.J. would like to thank UPCT for partial financial support through the concession of "Maria Zambrano ayudas para la recualificación del sistema universitario español 2021-2023" financed by Spanish Ministry of Universities with financial funds "Next Generation" of the EU.

\appendix

\section{Particle scattering by  array of periodic \texorpdfstring{$\mathcal{P}\mathcal{T}$}{PT}-symmetric cells}\label{scattgen} 
 
 Considering the scattering of a spinless particle of mass $m$ by assembly of periodic $\mathcal{P}\mathcal{T}$-symmetric cells, the   dynamics is described by one dimensional Schr\"odinger equation along incident direction $x$,
 \begin{equation}
 \left [ - \frac{1}{2 m} \frac{d^2}{d x^2} + \sum_{n= - N}^N V(x- n L) \right ] \Psi_E(x) =  E\Psi_E(x) ,
 \end{equation}
  where $V(x)$ stands for the potential in a unit cell. Total $2 N+1$ cells are placed symmetrically on both sides of origin. The length of scattering barriers is thus $(2 N+1) L$, where $L$ denotes the length of a single cell. Inside of each single cell,  we adopt a simple $\mathcal{P}\mathcal{T}$-symmetric impurities model with a potential:
  \begin{equation}
  V(x) = V \delta(x -a) + V^* \delta(x+a), \ \ \ \ V = |V| e^{i \theta}. \label{VPTpot}
  \end{equation}
  Two  contact interactions with the complex strength that are conjugate to each other are placed on two sides of cell's center with equal distance $a$, thus potential satisfies $\mathcal{P}\mathcal{T}$ symmetry relation:
  \begin{equation}
  V(-x) = V^*(x).
  \end{equation}

  For scattering solutions, it is more convenient to consider Lippmann-Schwinger (LS) equation,
  \begin{align}
 &  \Psi_E(x)  =\Psi^{(0)}_E(x)   \nonumber \\
  &+  \sum_{n= - N}^N \int_{- \infty}^\infty d x' G_0(x-x';E) V(x'- n L) \Psi_E(x') ,
  \end{align}
  where  
  \begin{equation} 
  \Psi^{(0)}_E(x)   =A e^{i k x} +B e^{- i k x} \label{incidentwav}
  \end{equation}
   represents the incident waves with   linear superposition of both left and right propagating plane waves, and $$  k=\sqrt{2 m (E+ i 0 )}$$ stands for the linear momentum of incident particle.
The Green's function of free particle,  $G_0(x;E)$, is given by
  \begin{equation}
  G_0(x;E) = \int_{-\infty}^\infty \frac{d p}{2\pi} \frac{e^{i p x}}{E - \frac{p^2}{2m}} = - \frac{ i m }{k} e^{i k |x|}.
  \end{equation}
  With contact interactions potential given in Eq.(\ref{VPTpot}), the scattering dynamics is thus totally determined by discrete LS equations on scattering sites,
  \begin{equation}
   \sum_{n'= - N}^N \left [ D(k) \right ]_{n, n' } 
   \begin{bmatrix}
    \Psi_E( n' L + a) \\
     \Psi_E( n' L - a)
   \end{bmatrix} =  \begin{bmatrix}
    \Psi^{(0)}_E( n L + a) \\
     \Psi^{(0)}_E( n L - a)
   \end{bmatrix}  , \label{wavdissol}
  \end{equation}
  where
    \begin{align}
 &   \left [ D(k) \right ]_{n, n' }  \nonumber \\
 &  = 
   \begin{bmatrix}
  \delta_{n, n'} + \frac{i m V}{k} e^{i k | n-n'| L} &   \frac{i m V^* }{k} e^{i k | nL-n'L + 2 a| }   \\
    \frac{i m V }{k} e^{i k | n L-n'L - 2 a| } &   \delta_{n, n'} + \frac{i m V^*}{k} e^{i k | n-n'| L} 
   \end{bmatrix} . \label{Dmat}
  \end{align}
  The on-energy-shell scattering amplitudes can be easily pulled out by considering asymptotic form of wave function,
  \begin{equation}
   \Psi_E(x)   \stackrel{|x| > NL}{=}\Psi^{(0)}_E(x)   + i f_E ( p ) e^{i k |x|}, \ \  p = \frac{x}{|x|} k,
  \end{equation}
  where
     \begin{align}
 &  f_E( p ) = -  \sum_{n= - N}^N e^{- i p n L}   \nonumber \\
 & \times \left [  \frac{m V}{k}  e^{- i p a}    \Psi_E( n L + a) +  \frac{m V^*}{k}  e^{ i p a}    \Psi_E( n L -a)  \right ]. \label{scatampdef}
  \end{align}
 Using Eq.(\ref{wavdissol}), the formal solution of scattering amplitude with a general incident wave, $A e^{i k x} +B e^{- i k x} $, is thus given by
     \begin{align}
 &  f_E ( p ) = -  \sum_{n, n'= - N}^N e^{- i p n L}  \nonumber \\
 & \times   \begin{bmatrix}
     \frac{m V}{k}  e^{- i p a}   &  \frac{m V^*}{k}  e^{ i p a}  
   \end{bmatrix}     \left [ D^{-1}(k) \right ]_{n, n' } 
   \begin{bmatrix}
    \Psi^{(0)}_E( n' L + a) \\
     \Psi^{(0)}_E( n' L - a)
   \end{bmatrix}  .  
  \end{align}

 The $S$-matrix in left/right propagating wave basis, (R) $A=1$ and $ B=0$; (L) $A=0 $ and $ B=1$, is defined by 
 \begin{equation}
 S(E ) =    \begin{bmatrix}
     t(k) & r^{(L)} (k)\\
      r^{(R)} (k ) & t(k) 
   \end{bmatrix}   ,
 \end{equation}
  where $t(k)$ and $r^{(L/R)} (k)$ are transmission and left/right reflection amplitudes, and they are related to the left/right basis scattering amplitudes by
  \begin{align}
 &  t(k) = 1+ i f_E^{(R)} (k) =1+ i f_E^{(L)} (- k) , \nonumber \\
 &   r^{(R)} (k ) = i f_E^{(R)} (-k), \ \ \ \   r^{(L)} (k ) = i f_E^{(L)} (k). 
  \end{align}
   Transmission and  reflection amplitudes  can   be parameterized by three real functions: one inelasticity $\eta(k) \in [1, \infty]$ and two phaseshifts $\delta_{\pm} (k)$, see Ref.~\cite{PhysRevResearch.4.023083},
   \begin{align}
   t &= \eta  \cos ( \delta_+  - \delta_- ) e^{ i (  \delta_+   +\delta_-   ) }, \nonumber \\
    r^{(R/L)}  & = i \left [ \eta  \sin ( \delta_+   - \delta_-  )  \pm  \sqrt{\eta^2  - 1} \right ]e^{ i (  \delta_+   +\delta_-   ) }. \label{Smatparam}
   \end{align}
   The inelasticity and phaseshifts are linked to scattering amplitudes directly by relations, see Ref.~\cite{PhysRevResearch.4.023083},
   \begin{align}
  & \frac{\eta(k) e^{2 i \delta_\pm (k)} -1}{2i}  \nonumber \\
  &= \frac{ \left [ f^{(R)} (k) \pm f^{(L)} (k) \right ] \pm \left [ f^{(R)} (-k) \pm f^{(L)} (-k)  \right ] }{4}.
   \end{align}
 For a   $\mathcal{P}\mathcal{T}$-symmetric system, using relations given in Eq.(\ref{Smatparam}), one can easily verify that
 \begin{equation}
 \frac{1}{2 i} \ln \det \left [ S(E) \right ] = \delta_+ (k) + \delta_- (k) = Im \left [ \ln t(k) \right ].
 \end{equation}

In addition, another very useful relation between transmission amplitude and $D (k)= 1- G_0 (k) V$ matrix defined in Eq.(\ref{Dmat}) is given by
\begin{equation}
t(k) = \frac{1}{\det \left [ D(k) \right ]}. \label{tdetDrel}
\end{equation}
 This relation can be proven by using the properties of $S$-matrix,  the $S$-matrix operator is related to $D$-matrix operator by, see Eq.(12), (24) and (27) in Ref.~\cite{PhysRevResearch.4.023083},
 \begin{equation}
 \hat{S} (E) = \frac{\hat{D} (-k)}{\hat{D} (k)}, \ \ \ \  \pm k = \sqrt{2m (E\pm i 0)},
 \end{equation}
 where $\hat{D} ( \pm k)$ are defined below and above branch cut singularity of  the analytic $\hat{D}$-matrix operator  respectively, and the branch cut is sitting on real energy axis. Hence we obtain
    \begin{equation}
 \frac{1}{2 i} \ln \det \left [ S(E) \right ]   = -  Im \left [ \ln \det [ D(k) ] \right ] = Im \left [ \ln t(k) \right ] .
 \end{equation}
Both $t(k)$ and  $\det [ D(k) ]$ are analytic functions defined in entire complex $E$-plane,      real parts of two functions are  related to  imaginary parts  by Cauchy integral,  which ultimately yields  the relation in Eq.(\ref{tdetDrel}).

\subsection{Analytic scattering solutions by LS equation approach}\label{analyticsol}  
 For the simple contact interaction $\mathcal{P}\mathcal{T}$-symmetric model, it turns out that the scattering solutions can be obtained analytically. The discrete LS equations in  Eq.(\ref{wavdissol}) can be solved by making assumption: 
      \begin{align}
 &    \begin{bmatrix}
    \Psi_E( n L + a) \\
     \Psi_E( n L - a)
   \end{bmatrix}  =   \begin{bmatrix}
   C &  D   \\
   E  & F  
   \end{bmatrix}    
   \begin{bmatrix}
 \cos ( Q  n L)   \\
    i \sin ( Q   n L  )
   \end{bmatrix}  .   \label{wavassumption}
  \end{align}
 That is to say that the wave function at scattering sites are determined completely by   collective modes  of  entire lattice  of impurities  and hence are  described by plane waves with a wave vector $Q$. All the coefficients, $(C,D,E,F)$, and wave vector, $Q$, can be determined by plugging Eq.(\ref{wavassumption}) into Eq.(\ref{wavdissol}).
 Using identities
 \begin{align}
 &  \sum_{n'= - N}^N  e^{i k |nL - n' L + d |} \cos (Q n' L)  \nonumber \\
 & =i  \frac{  \sin  ( k |d| - k L) - \sin( k |d |) \cos (Q L)   }{\cos (k L) - \cos (QL)} \cos (Q n L) \nonumber \\
 & + i  \frac{   \sin ( k d)  \sin  (Q L)   }{\cos (k L) - \cos (QL)} \sin (Q n L) \nonumber \\
 &-\mathbb{C} (k)  \cos (k n L + k d),
 \end{align} 
 and
  \begin{align}
 &  \sum_{n'= - N}^N  e^{i k |nL - n' L + d |}  \sin (Q n' L)  \nonumber \\
 & =i  \frac{  \sin  ( k |d| - k L) - \sin( k |d |) \cos (Q L)   }{\cos (k L) - \cos (QL)} \sin (Q n L) \nonumber \\
 & - i  \frac{   \sin ( k d)  \sin  (Q L)   }{\cos (k L) - \cos (QL)} \cos (Q n L) \nonumber \\
 &+ i \mathbb{S} (k)  \sin (k n L + k d),
 \end{align} 
 where $d = \pm 2 a$, and
 \begin{align}
\mathbb{C} (k) &= \frac{ \left [ \cos (Q ( N+1) L) - \cos (Q NL) e^{i k L} \right ] e^{ i k NL} }{\cos (k L) - \cos (QL)} , \nonumber \\
\mathbb{S} (k) &= \frac{ \left [ \sin (Q ( N+1) L) - \sin (Q NL) e^{i k L} \right ] e^{ i k NL} }{\cos (k L) - \cos (QL)} ,
 \end{align}
 comparing both sides of Eq.(\ref{wavdissol}), all the coefficients of independent plane waves, $(\cos (Q n L), \sin (Qn L))$ and $(\cos (k n L), \sin (k n L))$, must all vanish. Hence, we find
 \begin{align}
 Q =&  \frac{1}{L}  \arccos \bigg [  \cos (kL) + \frac{m |V|}{k} 2 \cos \theta \sin (k L) \nonumber \\
 & + 2 ( \frac{m |V|}{k} )^2 \sin (k 2 a) \sin (k L - k 2 a)  \bigg ]. \label{QCPT}
 \end{align}
 The relation given in Eq.(\ref{QCPT}) is in fact the exact energy-momentum dispersion relation when $N \rightarrow \infty$ and  $\mathcal{P}\mathcal{T}$-symmetric system or a diatomic crystal system becomes totally periodic, see e.g. Refs.~\cite{AHMED2001231,GASPARIAN199772}. The wave vector $Q$ hence plays the role of  crystal-momentum, where crystal is formed by all the impurities placed at interaction sites. We also remark that although even for a finite system, the wave function at interaction sites is indeed periodic and simply given by Bloch waves, $\Psi_E( n L \pm a) \propto e^{\pm i Q n L}$, the entire  wave function as matter of fact is not periodic and doesn't satisfies Bloch theorem.
 
 The coefficients $C$ and $D$ are given by the solution of  coupled  algebra equations,
 \begin{align}
&\cos (k a)  \mathbb{C} (k)   \left [  (1+ \frac{  m V^*  }{k}   \alpha) C +  \frac{  m V^*  }{k}  \beta D \right ]  \nonumber \\
& + \sin (k a)  \mathbb{S} (k)   \left [  (1- \frac{  m V^*  }{k}   \alpha) D -  \frac{  m V^*  }{k}  \beta C \right ]  = - \frac{A+B}{\frac{  i m V  }{k} }, \nonumber \\
&\sin (k a)  \mathbb{C} (k)   \left [  (1- \frac{  m V^*  }{k}   \alpha) C -  \frac{  m V^*  }{k}  \beta D \right ]  \nonumber \\
& + \cos (k a)  \mathbb{S} (k)   \left [  (1+ \frac{  m V^*  }{k}   \alpha) D +  \frac{  m V^*  }{k}  \beta C \right ]  = - \frac{A-B}{\frac{   m V  }{k} },
 \end{align}
 where
  \begin{align}
& \alpha =     \frac{   \sin  ( k2 a - k L)    - \sin  (k  2a)  \cos (Q L )  }{   \cos (k L) - \cos (Q L) + \frac{ m V^*  }{k}   \sin (k  L)             }    ,  \nonumber \\
&  \beta  =  \frac{  i  \sin ( k 2a)    \sin (QL)   }{   \cos (k L) - \cos (Q L) + \frac{ m V^*  }{k}   \sin (k  L)            }   .
\end{align}
The coefficients $E$ and $F$ are related to $C$ and $D$  by
\begin{equation}
E =  \frac{  m V  }{k}   ( \alpha C +      \beta D) , \ \  F =  \frac{  m V  }{k}   (  \beta C +       \alpha D ).
\end{equation}

Using Eq.(\ref{wavassumption}) and Eq.(\ref{scatampdef}), the analytic expression of scattering amplitude is given by
     \begin{align}
 &  f_E( p ) \nonumber \\
 & = -   \frac{m V}{k}       \left [    \frac{m V^*}{k}  e^{ i p a}  (\alpha \Omega_c(p) +\beta  \Omega_s(p)  )+  e^{- i p a}   \Omega_c(p)      \right ] C \nonumber \\
  &    -   \frac{m V}{k}       \left [       \frac{m V^*}{k}  e^{ i p a}  ( \beta \Omega_c(p)  +      \alpha  \Omega_s(p)  )+     e^{- i p a}     \Omega_s(p)   \right ] D    ,
  \end{align}
  where
  \begin{align}
  & \Omega_c(p) =  \sum_{n= - N}^N e^{- i p n L}   \cos ( Q  n L)  \nonumber \\
    &= \frac{ \cos (k(N+1) L) \cos (QNL)- \cos (kN L) \cos (Q(N+1)L)}{\cos (p L) -\cos QL}, \nonumber \\
    & \Omega_s(p) = \sum_{n= - N}^N e^{- i p n L}   i \sin ( Q  n L)  \nonumber \\
    &= \frac{ \sin (k(N+1) L) \sin (QNL)- \sin (kN L) \sin (Q(N+1)L)}{\cos (p L) -\cos QL}.
  \end{align}
After some lengthy and highly non-trivial calculation, compact forms of transmission and reflection amplitudes can be found
\begin{align}
& \left [  t(k) e^{ i k (2N+1) L} \right ]^{-1} = \det [D(k) ] e^{- i k (2N+1) L} \nonumber \\
& = \cos (Q (2N+1) L) - i \sin (k L) \frac{\sin (Q(2N+1)L)}{\sin (QL)} \nonumber \\
& + 2  \frac{ i m |V|}{k}    \cos \theta \cos (k L)  \frac{\sin (Q(2N+1)L)}{\sin (QL)}  \nonumber \\
& + 2  i  ( \frac{m |V|}{k} )^2 \sin (k 2 a) \cos (k L - k 2 a)   \frac{\sin (Q(2N+1)L)}{\sin (QL)},
\end{align}
and
\begin{align}
r^{(R/L)} (k) &= -  \frac{\frac{i m |V|}{k} }{\det [D(k)]} \frac{\sin ( Q (2N+1) L)}{\sin (QL)}  \nonumber \\
& \times 2 \left [   \cos (    k2 a \pm \theta) +  \frac{ m |V|}{k}   \sin( k 2a)      \right ]  .
\end{align}

\subsection{Characteristic determinant approach}

The analytic solutions  can also be obtained  elegantly   by characteristic determinant approach that was developed in Refs.~\cite{GASPARIAN1988201,GASPARIAN199772,Aronov_1991}. The key idea is to take advantage of recursion relations that the  determinant of $D$-matrix   in Eq.(\ref{Dmat}) must satisfy. Starting with a single  impurity, by adding one impurity at a time and using recursion relations, and then expression of   determinant of $D$-matrix of a finite size multiple cells system can be obtained. 

The general idea of characteristic determinant approach  can be summarized as follows: considering a simple impurities model with potential of
\begin{equation}
V(x) = \sum_{n =1}^M V_n \delta (x - x_n), \ \ \ \ x_{n-1} < x_n,
\end{equation}
where  $V_n$ is the strength of contact interaction,  $x_n$ stands for the position of  the $n$-th  impurity  and  $M$ is the total numbers of  scatters. The matrix elements of $D $-matrix, $\hat{D}= 1- G_0 \hat{V}$, for this simple contact interaction model is thus given by
    \begin{equation}
  \left [ D(k) \right ]_{n, n' }  = 
  \delta_{n, n'} + \frac{i m V_n}{k} e^{i k | x_n- x_{n'}| } . 
  \end{equation}
Let's introduce a short-hand notation $$ \mathcal{D}^{(M)} = \det \left [ D(k) \right ]_{M \times M}$$ to  denote the determinant of $D$-matrix for a system with $M$ cells. The  determinant of $D$-matrix for   systems with $M$, $M-1$ and $M-2$ number of cells respectively are thus related by recursion relation
\begin{equation}
 \mathcal{D}^{(M)}  =\mathcal{A}^{(M)}  \mathcal{D}^{(M-1)}  -\mathcal{ B}^{(M)} \mathcal{D}^{(M-2)} ,
\end{equation}
 where
 \begin{align}
  \mathcal{ B}^{(M)}&  = \frac{V_M}{V_{M-1}} e^{ i k 2 (x_M -x_{M-1})}, \nonumber \\
    \mathcal{ A}^{(M)} & =1+  \mathcal{ B}^{(M)} +  \frac{i V_M}{ k } (1 - e^{ i k 2 (x_M -x_{M-1})} ).
\end{align}
The initial conditions for the recurrence relations are
\begin{equation}
 \mathcal{D}^{(-1)} =1, \ \ \ \  \mathcal{D}^{(0)} =1, \ \ \ \  \mathcal{A}^{(1)} =1 + \frac{i m V_1}{k} = \mathcal{D}^{(1)}.
\end{equation}

 The transmission amplitude for a  $M$ cells  system is thus simply given by $t (k) =1/ \mathcal{D}^{(M)} (k)$. Once the transmission amplitude is given, for contact interactions model, the reflection amplitudes may be worked out simply by matching boundary condition at site of each scatter. The reflection and transmission amplitudes are thus related by
\begin{align}
&  \begin{bmatrix} 
  1 \\ 
  r^{(R)} (k)
  \end{bmatrix}  = \mathcal{M}^{(M)} (k )   \begin{bmatrix} 
  t (k)   \\ 
  0  
  \end{bmatrix}  , \nonumber \\
  &   \begin{bmatrix} 
 0   \\ 
   t (k) 
  \end{bmatrix}  = \mathcal{M}^{(M)} (k )   \begin{bmatrix} 
  r^{(L)} (k) \\ 
  1
  \end{bmatrix}  ,
\end{align} 
where the transfer matrix for a $M$ cells  system is given by
\begin{equation}
 \mathcal{M}^{(M)} (k ) 
  =    \prod_{n=1}^M   \begin{bmatrix} 
 1+ \frac{i m V_n}{k} &  \frac{i m V_n }{k}e^{-2  i k x_n }    \\ 
 - \frac{i m V_n }{k}     e^{2i k x_n}  &1 -  \frac{i m V_n }{k}
  \end{bmatrix} ,
 \end{equation}
 and 
 \begin{equation}
 \det \left [  \mathcal{M}^{(M)} (k )\right ] =1.
 \end{equation}
The transfer matrix can   be parameterized in terms of transmission and reflection amplitudes by
 \begin{equation}
 \mathcal{M}^{(M)} (k ) 
  =       \begin{bmatrix} 
\frac{1}{ t(k ) }&   - \frac{r^{(L)} (k)}{t(k)}     \\ 
 \frac{r^{(R)} (k)}{t(k)}&  t(k )-  \frac{r^{(L)} (k) r^{(R)} (k)}{t(k)}
  \end{bmatrix} . \label{transfermat}
 \end{equation}
 The left propagating reflection amplitude, $r^{(L)} (k)$, can be obtained by  taking advantage of relation of transfer matrix
  \begin{equation}
  \mathcal{M}^{(M)} (k ) =  \mathcal{M}^{(M-1)} (k )  \begin{bmatrix} 
 1+ \frac{i m V_M }{k} &  \frac{i m V_M }{k}e^{-2  i k x_M }    \\ 
 - \frac{i m V_M }{k}     e^{2i k x_M}  &1 -  \frac{i m V_M }{k}
  \end{bmatrix} ,
 \end{equation}
 which describes $M$ cells system is composed of $(M-1) $ cells counting from left to right plus $M$-th cell sitting on the right edge of system. Using Eq.(\ref{transfermat}) for both $M$ cells and $(M-1)$ cells systems, thus  the  left propagating reflection amplitude, $r^{(L)} (k)$, for a $M$ cells system  is related to determinant of $D$-matrix by
  \begin{equation}
     r^{(L)} (k)     = \left [   \frac{  1     -     \frac{ \mathcal{D}^{(M-1)} (k)}{ \mathcal{D}^{(M)} ( k)} }{\frac{i m V_M }{k}  } -1 \right ]  e^{ - 2 i k x_M } .
\end{equation}
The  right propagating reflection amplitude $r^{(R)} (k)$ can be obtained  by the same procedure by reversing the direction of operation and counting from right to left.   For the $\mathcal{P}\mathcal{T}$-symmetric system, left/right reflection amplitudes are related by symmetry relation, see Ref.~\cite{PhysRevResearch.4.023083},
   \begin{equation}
     r^{(L)} (- k)     =  r^{(R) *} (k)  .
\end{equation}

With some length calculation, finally for   the diatomic periodic $\mathcal{P}\mathcal{T}$-symmetric model, we find again
  \begin{align}
&  t(k)    = \frac{1}{\det [D(k)]} = \frac{ \sec (Q(2N+1)  L ) e^{ - i k (2N+1) L} }{   1+ i  Im \left [ \frac{e^{- i k L} }{t_0(k)}    \right ] \frac{\tan  (Q (2N+1) L)}{\sin (Q L)}    },  \nonumber \\
&   \frac{ r^{(L/R)} (k) }{t(k)} =   \left [  \frac{ r^{(L/R)}_0 (k) }{t_0(k)}  \right ] \frac{\sin (Q (2N+1) L)}{\sin (QL) }     ,  
\end{align}
where  $Q$  is defined in Eq.(\ref{QCPT}).  The transmission and reflection amplitudes by a single cell,   $t_0 (k)$  and $ r^{(L/R)}_0 (k)$ respectively,  are    given by
\begin{align}
& \frac{1 }{t_0(k)}  =  1+ 2 \frac{ i m  | V|  }{k  } \cos \theta   + 2 i  \left  ( \frac{  m|V| }{k} \right  )^2 \sin ( k 2 a) e^{  i  k 2a    }    , \nonumber \\
 & \frac{ r^{(L/R)}_0 (k) }{t_0(k)}   =  - 2 \frac{i m |V|}{k}    \left [   \cos (    k2 a \mp \theta) +  \frac{ m |V|}{k}   \sin (k 2a)      \right ] .
\end{align}
The result for a general diatomic model can be found in e.g. Refs.~\cite{GASPARIAN199772,Aronov_1991}.

\section{Periodic \texorpdfstring{$\mathcal{P}\mathcal{T}$}{PT}-symmetric systems}\label{finitevolumesol}
Let's also consider a periodic system 
 \begin{equation}
 \left [ - \frac{1}{2 m} \frac{d^2}{d x^2} +  V_L(x ) \right ] \Psi^{(Q,L)}_E(x) =  E\Psi^{(Q,L)}_E(x) ,
 \end{equation}
  where $$V_L(x  ) = \sum_{n = - \infty}^{\infty} V(x+ n L)$$ is a periodic potential:  $V_L(x + n L) = V_L(x)$. The wave function satisfies periodic boundary condition,
  \begin{equation}
  \Psi^{(Q,L)}_E(x+ n L) = e^{ i Q n L} \Psi^{(Q,L)}_E(x) ,
  \end{equation}
where superscript $Q$ is added to label the $Q$-dependence of periodic boundary condition. The stationary solutions can be found by homogeneous LS equation for a single cell, see Refs.~\cite{GUO2020135370,PhysRevD.88.014507,PhysRevD.88.014501,PhysRevD.95.054508,PhysRevD.101.094510},
 \begin{equation}
  \Psi^{(Q,L)}_E(x) =   \int_{- \frac{L}{2}}^{\frac{L}{2}} d x' G^{(Q,L)}_0(x-x';E) V_L(x') \Psi^{(Q,L)}_E(x') ,
  \end{equation}
  where  periodic Green's function of a free particle,  $G^{(Q,L)}_0(x;E)$, is  defined by
  \begin{align}
  &G^{(Q, L)}_0(x;E)  =\sum_{n= - \infty}^\infty G_0(x+ n L;E) e^{ -i Q n L}   \nonumber \\
  & = \frac{1}{L} \sum_{p = \frac{2\pi n}{L} + Q, n \in \mathbb{Z}} \frac{e^{i p x}}{E - \frac{p^2}{2m}}  \nonumber \\
  &= - \frac{ i m }{k}  \left [ e^{i k |x|} + \frac{\cos (k x- QL) - \cos (k x) e^{ i k L}}{\cos (k L) - \cos (QL)} \right ].
  \end{align}
With contact interactions potential in Eq.(\ref{VPTpot}), the quantization condition for eigensolutions is given by
  \begin{equation}
  \det
\begin{bmatrix}
1- V  G^{(Q,L)}_0(0;E) &  - V^*  G^{(Q,L)}_0(2a;E)  \\
 - V G^{(Q,L)}_0(-2 a;E) &1- V^*  G^{(Q,L)}_0(0;E)  
\end{bmatrix} =0.
\end{equation}
Hence we again get well-known energy-momentum dispersion relation for a periodic system, see e.g. Refs.~\cite{AHMED2001231,GASPARIAN199772}, 
 \begin{align}
 \cos (Q L) &=     \cos (kL) + \frac{m |V|}{k} 2 \cos \theta \sin (k L) \nonumber \\
 & + 2 ( \frac{m |V|}{k} )^2 \sin (k 2 a) \sin (k L - k 2 a)   .  
 \end{align}

For a periodic  $\mathcal{P}\mathcal{T}$-symmetric system, a generalized    density of  states for a single cell may be defined by
\begin{equation}
n^{(Q,L)}_E (x)=-\frac{1}{\pi}  Im  \left [ \langle  x | \hat{G}^{(Q, L)} (E) |x \rangle  \right ].
\end{equation}
The spectral representation of full Green's function operator is given by, see e.g. Appendix C in Ref.~\cite{PhysRevResearch.4.023083},
\begin{equation}
\hat{G}^{(Q, L)} (E) = \sum_{i} \frac{  |   \Psi_{E_i }^{(Q, L)}  \rangle \langle \widetilde{\Psi}_{E_i }^{(Q, L)}|}{E-E_i },
\end{equation}
where    $E_i = E_i (Q)$ is i-th band eigenvalue as the function of crystal-momentum $Q$, and  the sum is over all the allowed bands.
The $| \widetilde{\Psi}_E^{(Q,L)} \rangle$ represents the eigenstate of adjoint Hamiltonian $\hat{H}^\dag$,
\begin{equation}
\hat{H}^\dag  | \widetilde{\Psi}_E^{(Q, L)} \rangle = E | \widetilde{\Psi}_E^{(Q,L)} \rangle.
\end{equation}
The wave functions, $| \Psi_E^{(Q,L)} \rangle $ and  $ | \widetilde{\Psi}_E^{(Q,L)} \rangle  $, together are biorthogonal and can be normalized in a unit cell,
\begin{equation}
 \int_{-\frac{L}{2}}^{\frac{L}{2}} d x \left [ \langle  x   |   \Psi_{E_i }^{(Q, L)}  \rangle \langle \widetilde{\Psi}_{E_i }^{(Q, L)}|  x \rangle \right ] = 1.
\end{equation}
Hence the integrated  generalized    density of state for a periodic system is now given by
\begin{equation}
 \int_{BZ} d Q   \int_{-\frac{L}{2}}^{\frac{L}{2}} d x n^{(Q,L)}_E (x)   = \sum_{i}  \int_{BZ} d Q  \delta (E-E_i )   = \frac{d Q}{d E}   ,
\end{equation}
where    integration of the crystal-momentum is confined within the first Brillouin zone. We   remark that the above relation is only defined in allowed bands, the density of state should be defined as zero in gaps.

\bibliography{ALL-REF.bib}

\begin{thebibliography}{74}
\expandafter\ifx\csname natexlab\endcsname\relax\def\natexlab#1{#1}\fi
\expandafter\ifx\csname bibnamefont\endcsname\relax
  \def\bibnamefont#1{#1}\fi
\expandafter\ifx\csname bibfnamefont\endcsname\relax
  \def\bibfnamefont#1{#1}\fi
\expandafter\ifx\csname citenamefont\endcsname\relax
  \def\citenamefont#1{#1}\fi
\expandafter\ifx\csname url\endcsname\relax
  \def\url#1{\texttt{#1}}\fi
\expandafter\ifx\csname urlprefix\endcsname\relax\def\urlprefix{URL }\fi
\providecommand{\bibinfo}[2]{#2}
\providecommand{\eprint}[2][]{\url{#2}}

\bibitem[{\citenamefont{Moiseyev}(2011)}]{moiseyev_2011}
\bibinfo{author}{\bibfnamefont{N.}~\bibnamefont{Moiseyev}},
  \emph{\bibinfo{title}{Non-Hermitian Quantum Mechanics}}
  (\bibinfo{publisher}{Cambridge University Press}, \bibinfo{year}{2011}).

\bibitem[{\citenamefont{Landauer and Martin}(1994)}]{RevModPhys.66.217}
\bibinfo{author}{\bibfnamefont{R.}~\bibnamefont{Landauer}} \bibnamefont{and}
  \bibinfo{author}{\bibfnamefont{T.}~\bibnamefont{Martin}},
  \bibinfo{journal}{Rev. Mod. Phys.} \textbf{\bibinfo{volume}{66}},
  \bibinfo{pages}{217} (\bibinfo{year}{1994}),
  \urlprefix\url{https://link.aps.org/doi/10.1103/RevModPhys.66.217}.

\bibitem[{\citenamefont{Gasparian et~al.}(2000)\citenamefont{Gasparian,
  Ortuño, Schön, and Simon}}]{GASPARIAN2000513}
\bibinfo{author}{\bibfnamefont{V.}~\bibnamefont{Gasparian}},
  \bibinfo{author}{\bibfnamefont{M.}~\bibnamefont{Ortuño}},
  \bibinfo{author}{\bibfnamefont{G.}~\bibnamefont{Schön}}, \bibnamefont{and}
  \bibinfo{author}{\bibfnamefont{U.}~\bibnamefont{Simon}}, in
  \emph{\bibinfo{booktitle}{Handbook of Nanostructured Materials and
  Nanotechnology}}, edited by \bibinfo{editor}{\bibfnamefont{H.~S.}
  \bibnamefont{Nalwa}} (\bibinfo{publisher}{Academic Press},
  \bibinfo{address}{Burlington}, \bibinfo{year}{2000}), pp.
  \bibinfo{pages}{513--569}, ISBN \bibinfo{isbn}{978-0-12-513760-7},
  \urlprefix\url{https://www.sciencedirect.com/science/article/pii/B9780125137607500277}.

\bibitem[{\citenamefont{Fayer and Fayer}(2001)}]{fayer2001elements}
\bibinfo{author}{\bibfnamefont{M.}~\bibnamefont{Fayer}} \bibnamefont{and}
  \bibinfo{author}{\bibfnamefont{P.}~\bibnamefont{Fayer}},
  \emph{\bibinfo{title}{Elements of Quantum Mechanics}}
  (\bibinfo{publisher}{Oxford University Press}, \bibinfo{year}{2001}), ISBN
  \bibinfo{isbn}{9780195141955},
  \urlprefix\url{https://books.google.com/books?id=Lf11rpyQMXwC}.

\bibitem[{\citenamefont{B{\"u}ttiker et~al.}(1994)\citenamefont{B{\"u}ttiker,
  Thomas, and Pr{\^e}tre}}]{Buttiker1994}
\bibinfo{author}{\bibfnamefont{M.}~\bibnamefont{B{\"u}ttiker}},
  \bibinfo{author}{\bibfnamefont{H.}~\bibnamefont{Thomas}}, \bibnamefont{and}
  \bibinfo{author}{\bibfnamefont{A.}~\bibnamefont{Pr{\^e}tre}},
  \bibinfo{journal}{Zeitschrift f{\"u}r Physik B Condensed Matter}
  \textbf{\bibinfo{volume}{94}}, \bibinfo{pages}{133} (\bibinfo{year}{1994}),
  \urlprefix\url{https://doi.org/10.1007/BF01307664}.

\bibitem[{\citenamefont{Brouwer}(1998)}]{PhysRevB.58.R10135}
\bibinfo{author}{\bibfnamefont{P.~W.} \bibnamefont{Brouwer}},
  \bibinfo{journal}{Phys. Rev. B} \textbf{\bibinfo{volume}{58}},
  \bibinfo{pages}{R10135} (\bibinfo{year}{1998}),
  \urlprefix\url{https://link.aps.org/doi/10.1103/PhysRevB.58.R10135}.

\bibitem[{\citenamefont{Zhou et~al.}(1999)\citenamefont{Zhou, Spivak, and
  Altshuler}}]{PhysRevLett.82.608}
\bibinfo{author}{\bibfnamefont{F.}~\bibnamefont{Zhou}},
  \bibinfo{author}{\bibfnamefont{B.}~\bibnamefont{Spivak}}, \bibnamefont{and}
  \bibinfo{author}{\bibfnamefont{B.}~\bibnamefont{Altshuler}},
  \bibinfo{journal}{Phys. Rev. Lett.} \textbf{\bibinfo{volume}{82}},
  \bibinfo{pages}{608} (\bibinfo{year}{1999}),
  \urlprefix\url{https://link.aps.org/doi/10.1103/PhysRevLett.82.608}.

\bibitem[{\citenamefont{Leavens and Aers}(1987)}]{LEAVENS19871101}
\bibinfo{author}{\bibfnamefont{C.}~\bibnamefont{Leavens}} \bibnamefont{and}
  \bibinfo{author}{\bibfnamefont{G.}~\bibnamefont{Aers}},
  \bibinfo{journal}{Solid State Communications} \textbf{\bibinfo{volume}{63}},
  \bibinfo{pages}{1101} (\bibinfo{year}{1987}), ISSN \bibinfo{issn}{0038-1098},
  \urlprefix\url{https://www.sciencedirect.com/science/article/pii/003810988791057X}.

\bibitem[{\citenamefont{Büttiker and Landauer}(1985)}]{MButtiker_1985}
\bibinfo{author}{\bibfnamefont{M.}~\bibnamefont{Büttiker}} \bibnamefont{and}
  \bibinfo{author}{\bibfnamefont{R.}~\bibnamefont{Landauer}},
  \bibinfo{journal}{Physica Scripta} \textbf{\bibinfo{volume}{32}},
  \bibinfo{pages}{429} (\bibinfo{year}{1985}),
  \urlprefix\url{https://dx.doi.org/10.1088/0031-8949/32/4/031}.

\bibitem[{\citenamefont{Buttiker and Landauer}(1986)}]{Buttiker5390141}
\bibinfo{author}{\bibfnamefont{M.}~\bibnamefont{Buttiker}} \bibnamefont{and}
  \bibinfo{author}{\bibfnamefont{R.}~\bibnamefont{Landauer}},
  \bibinfo{journal}{IBM Journal of Research and Development}
  \textbf{\bibinfo{volume}{30}}, \bibinfo{pages}{451} (\bibinfo{year}{1986}).

\bibitem[{\citenamefont{Martin and Landauer}(1992)}]{PhysRevB.45.1742}
\bibinfo{author}{\bibfnamefont{T.}~\bibnamefont{Martin}} \bibnamefont{and}
  \bibinfo{author}{\bibfnamefont{R.}~\bibnamefont{Landauer}},
  \bibinfo{journal}{Phys. Rev. B} \textbf{\bibinfo{volume}{45}},
  \bibinfo{pages}{1742} (\bibinfo{year}{1992}),
  \urlprefix\url{https://link.aps.org/doi/10.1103/PhysRevB.45.1742}.

\bibitem[{\citenamefont{B\"uttiker and Landauer}(1982)}]{PhysRevLett.49.1739}
\bibinfo{author}{\bibfnamefont{M.}~\bibnamefont{B\"uttiker}} \bibnamefont{and}
  \bibinfo{author}{\bibfnamefont{R.}~\bibnamefont{Landauer}},
  \bibinfo{journal}{Phys. Rev. Lett.} \textbf{\bibinfo{volume}{49}},
  \bibinfo{pages}{1739} (\bibinfo{year}{1982}),
  \urlprefix\url{https://link.aps.org/doi/10.1103/PhysRevLett.49.1739}.

\bibitem[{\citenamefont{Landauer}(1993)}]{LandauerNature1993}
\bibinfo{author}{\bibfnamefont{R.}~\bibnamefont{Landauer}},
  \bibinfo{journal}{Nature} \textbf{\bibinfo{volume}{365}},
  \bibinfo{pages}{692} (\bibinfo{year}{1993}),
  \urlprefix\url{https://doi.org/10.1038/365692a0}.

\bibitem[{\citenamefont{Japha and Kurizki}(1999)}]{PhysRevA.60.1811}
\bibinfo{author}{\bibfnamefont{Y.}~\bibnamefont{Japha}} \bibnamefont{and}
  \bibinfo{author}{\bibfnamefont{G.}~\bibnamefont{Kurizki}},
  \bibinfo{journal}{Phys. Rev. A} \textbf{\bibinfo{volume}{60}},
  \bibinfo{pages}{1811} (\bibinfo{year}{1999}),
  \urlprefix\url{https://link.aps.org/doi/10.1103/PhysRevA.60.1811}.

\bibitem[{\citenamefont{Muga et~al.}(1992)\citenamefont{Muga, Brouard, and
  Sala}}]{MUGA199224}
\bibinfo{author}{\bibfnamefont{J.}~\bibnamefont{Muga}},
  \bibinfo{author}{\bibfnamefont{S.}~\bibnamefont{Brouard}}, \bibnamefont{and}
  \bibinfo{author}{\bibfnamefont{R.}~\bibnamefont{Sala}},
  \bibinfo{journal}{Physics Letters A} \textbf{\bibinfo{volume}{167}},
  \bibinfo{pages}{24} (\bibinfo{year}{1992}), ISSN \bibinfo{issn}{0375-9601},
  \urlprefix\url{https://www.sciencedirect.com/science/article/pii/0375960192906202}.

\bibitem[{\citenamefont{B\"uttiker}(1983)}]{PhysRevB.27.6178}
\bibinfo{author}{\bibfnamefont{M.}~\bibnamefont{B\"uttiker}},
  \bibinfo{journal}{Phys. Rev. B} \textbf{\bibinfo{volume}{27}},
  \bibinfo{pages}{6178} (\bibinfo{year}{1983}),
  \urlprefix\url{https://link.aps.org/doi/10.1103/PhysRevB.27.6178}.

\bibitem[{\citenamefont{Sokolovski and Baskin}(1987)}]{PhysRevA.36.4604}
\bibinfo{author}{\bibfnamefont{D.}~\bibnamefont{Sokolovski}} \bibnamefont{and}
  \bibinfo{author}{\bibfnamefont{L.~M.} \bibnamefont{Baskin}},
  \bibinfo{journal}{Phys. Rev. A} \textbf{\bibinfo{volume}{36}},
  \bibinfo{pages}{4604} (\bibinfo{year}{1987}),
  \urlprefix\url{https://link.aps.org/doi/10.1103/PhysRevA.36.4604}.

\bibitem[{\citenamefont{Gasparian and Pollak}(1993)}]{PhysRevB.47.2038}
\bibinfo{author}{\bibfnamefont{V.}~\bibnamefont{Gasparian}} \bibnamefont{and}
  \bibinfo{author}{\bibfnamefont{M.}~\bibnamefont{Pollak}},
  \bibinfo{journal}{Phys. Rev. B} \textbf{\bibinfo{volume}{47}},
  \bibinfo{pages}{2038} (\bibinfo{year}{1993}),
  \urlprefix\url{https://link.aps.org/doi/10.1103/PhysRevB.47.2038}.

\bibitem[{\citenamefont{Gasparian
  et~al.}(1995{\natexlab{a}})\citenamefont{Gasparian, Ortu\~no, Ruiz, Cuevas,
  and Pollak}}]{PhysRevB.51.6743}
\bibinfo{author}{\bibfnamefont{V.}~\bibnamefont{Gasparian}},
  \bibinfo{author}{\bibfnamefont{M.}~\bibnamefont{Ortu\~no}},
  \bibinfo{author}{\bibfnamefont{J.}~\bibnamefont{Ruiz}},
  \bibinfo{author}{\bibfnamefont{E.}~\bibnamefont{Cuevas}}, \bibnamefont{and}
  \bibinfo{author}{\bibfnamefont{M.}~\bibnamefont{Pollak}},
  \bibinfo{journal}{Phys. Rev. B} \textbf{\bibinfo{volume}{51}},
  \bibinfo{pages}{6743} (\bibinfo{year}{1995}{\natexlab{a}}),
  \urlprefix\url{https://link.aps.org/doi/10.1103/PhysRevB.51.6743}.

\bibitem[{\citenamefont{Gasparian et~al.}(1996)\citenamefont{Gasparian,
  Christen, and B\"uttiker}}]{PhysRevA.54.4022}
\bibinfo{author}{\bibfnamefont{V.}~\bibnamefont{Gasparian}},
  \bibinfo{author}{\bibfnamefont{T.}~\bibnamefont{Christen}}, \bibnamefont{and}
  \bibinfo{author}{\bibfnamefont{M.}~\bibnamefont{B\"uttiker}},
  \bibinfo{journal}{Phys. Rev. A} \textbf{\bibinfo{volume}{54}},
  \bibinfo{pages}{4022} (\bibinfo{year}{1996}),
  \urlprefix\url{https://link.aps.org/doi/10.1103/PhysRevA.54.4022}.

\bibitem[{\citenamefont{Gasparian
  et~al.}(1995{\natexlab{b}})\citenamefont{Gasparian, Ortu\~no, Ruiz, and
  Cuevas}}]{PhysRevLett.75.2312}
\bibinfo{author}{\bibfnamefont{V.}~\bibnamefont{Gasparian}},
  \bibinfo{author}{\bibfnamefont{M.}~\bibnamefont{Ortu\~no}},
  \bibinfo{author}{\bibfnamefont{J.}~\bibnamefont{Ruiz}}, \bibnamefont{and}
  \bibinfo{author}{\bibfnamefont{E.}~\bibnamefont{Cuevas}},
  \bibinfo{journal}{Phys. Rev. Lett.} \textbf{\bibinfo{volume}{75}},
  \bibinfo{pages}{2312} (\bibinfo{year}{1995}{\natexlab{b}}),
  \urlprefix\url{https://link.aps.org/doi/10.1103/PhysRevLett.75.2312}.

\bibitem[{\citenamefont{Gasparian et~al.}(1999)\citenamefont{Gasparian,
  Sch{\"o}n, Ruiz, and Ortu{\~n}o}}]{Gasparian1999}
\bibinfo{author}{\bibfnamefont{V.}~\bibnamefont{Gasparian}},
  \bibinfo{author}{\bibfnamefont{G.}~\bibnamefont{Sch{\"o}n}},
  \bibinfo{author}{\bibfnamefont{J.}~\bibnamefont{Ruiz}}, \bibnamefont{and}
  \bibinfo{author}{\bibfnamefont{M.}~\bibnamefont{Ortu{\~n}o}},
  \bibinfo{journal}{The European Physical Journal B - Condensed Matter and
  Complex Systems} \textbf{\bibinfo{volume}{9}}, \bibinfo{pages}{283}
  (\bibinfo{year}{1999}),
  \urlprefix\url{https://doi.org/10.1007/s100510050767}.

\bibitem[{\citenamefont{Balcou and Dutriaux}(1997)}]{PhysRevLett.78.851}
\bibinfo{author}{\bibfnamefont{P.}~\bibnamefont{Balcou}} \bibnamefont{and}
  \bibinfo{author}{\bibfnamefont{L.}~\bibnamefont{Dutriaux}},
  \bibinfo{journal}{Phys. Rev. Lett.} \textbf{\bibinfo{volume}{78}},
  \bibinfo{pages}{851} (\bibinfo{year}{1997}),
  \urlprefix\url{https://link.aps.org/doi/10.1103/PhysRevLett.78.851}.

\bibitem[{\citenamefont{Muga et~al.}(2004)\citenamefont{Muga, Palao, Navarro,
  and Egusquiza}}]{MUGA2004357}
\bibinfo{author}{\bibfnamefont{J.}~\bibnamefont{Muga}},
  \bibinfo{author}{\bibfnamefont{J.}~\bibnamefont{Palao}},
  \bibinfo{author}{\bibfnamefont{B.}~\bibnamefont{Navarro}}, \bibnamefont{and}
  \bibinfo{author}{\bibfnamefont{I.}~\bibnamefont{Egusquiza}},
  \bibinfo{journal}{Physics Reports} \textbf{\bibinfo{volume}{395}},
  \bibinfo{pages}{357} (\bibinfo{year}{2004}), ISSN \bibinfo{issn}{0370-1573},
  \urlprefix\url{https://www.sciencedirect.com/science/article/pii/S0370157304001218}.

\bibitem[{\citenamefont{Hasan et~al.}(2020)\citenamefont{Hasan, Singh, and
  Mandal}}]{Hasan2020}
\bibinfo{author}{\bibfnamefont{M.}~\bibnamefont{Hasan}},
  \bibinfo{author}{\bibfnamefont{V.~N.} \bibnamefont{Singh}}, \bibnamefont{and}
  \bibinfo{author}{\bibfnamefont{B.~P.} \bibnamefont{Mandal}},
  \bibinfo{journal}{The European Physical Journal Plus}
  \textbf{\bibinfo{volume}{135}}, \bibinfo{pages}{640} (\bibinfo{year}{2020}),
  \urlprefix\url{https://doi.org/10.1140/epjp/s13360-020-00664-6}.

\bibitem[{\citenamefont{Jian et~al.}(2020)\citenamefont{Jian, Liu, Bai, Zhang,
  Zhang, Zhang, Xue, Sang, and Zhang}}]{JIAN2020125815}
\bibinfo{author}{\bibfnamefont{A.}~\bibnamefont{Jian}},
  \bibinfo{author}{\bibfnamefont{F.}~\bibnamefont{Liu}},
  \bibinfo{author}{\bibfnamefont{G.}~\bibnamefont{Bai}},
  \bibinfo{author}{\bibfnamefont{B.}~\bibnamefont{Zhang}},
  \bibinfo{author}{\bibfnamefont{Y.}~\bibnamefont{Zhang}},
  \bibinfo{author}{\bibfnamefont{Q.}~\bibnamefont{Zhang}},
  \bibinfo{author}{\bibfnamefont{X.}~\bibnamefont{Xue}},
  \bibinfo{author}{\bibfnamefont{S.}~\bibnamefont{Sang}}, \bibnamefont{and}
  \bibinfo{author}{\bibfnamefont{X.}~\bibnamefont{Zhang}},
  \bibinfo{journal}{Optics Communications} \textbf{\bibinfo{volume}{475}},
  \bibinfo{pages}{125815} (\bibinfo{year}{2020}), ISSN
  \bibinfo{issn}{0030-4018},
  \urlprefix\url{https://www.sciencedirect.com/science/article/pii/S0030401820303205}.

\bibitem[{\citenamefont{Bendix et~al.}(2009)\citenamefont{Bendix, Fleischmann,
  Kottos, and Shapiro}}]{PhysRevLett.103.030402}
\bibinfo{author}{\bibfnamefont{O.}~\bibnamefont{Bendix}},
  \bibinfo{author}{\bibfnamefont{R.}~\bibnamefont{Fleischmann}},
  \bibinfo{author}{\bibfnamefont{T.}~\bibnamefont{Kottos}}, \bibnamefont{and}
  \bibinfo{author}{\bibfnamefont{B.}~\bibnamefont{Shapiro}},
  \bibinfo{journal}{Phys. Rev. Lett.} \textbf{\bibinfo{volume}{103}},
  \bibinfo{pages}{030402} (\bibinfo{year}{2009}),
  \urlprefix\url{https://link.aps.org/doi/10.1103/PhysRevLett.103.030402}.

\bibitem[{\citenamefont{Schindler et~al.}(2011)\citenamefont{Schindler, Li,
  Zheng, Ellis, and Kottos}}]{PhysRevA.84.040101}
\bibinfo{author}{\bibfnamefont{J.}~\bibnamefont{Schindler}},
  \bibinfo{author}{\bibfnamefont{A.}~\bibnamefont{Li}},
  \bibinfo{author}{\bibfnamefont{M.~C.} \bibnamefont{Zheng}},
  \bibinfo{author}{\bibfnamefont{F.~M.} \bibnamefont{Ellis}}, \bibnamefont{and}
  \bibinfo{author}{\bibfnamefont{T.}~\bibnamefont{Kottos}},
  \bibinfo{journal}{Phys. Rev. A} \textbf{\bibinfo{volume}{84}},
  \bibinfo{pages}{040101} (\bibinfo{year}{2011}),
  \urlprefix\url{https://link.aps.org/doi/10.1103/PhysRevA.84.040101}.

\bibitem[{\citenamefont{Bender et~al.}(2013{\natexlab{a}})\citenamefont{Bender,
  Gianfreda, \"Ozdemir, Peng, and Yang}}]{PhysRevA.88.062111}
\bibinfo{author}{\bibfnamefont{C.~M.} \bibnamefont{Bender}},
  \bibinfo{author}{\bibfnamefont{M.}~\bibnamefont{Gianfreda}},
  \bibinfo{author}{\bibfnamefont{i.~m. c.~K.} \bibnamefont{\"Ozdemir}},
  \bibinfo{author}{\bibfnamefont{B.}~\bibnamefont{Peng}}, \bibnamefont{and}
  \bibinfo{author}{\bibfnamefont{L.}~\bibnamefont{Yang}},
  \bibinfo{journal}{Phys. Rev. A} \textbf{\bibinfo{volume}{88}},
  \bibinfo{pages}{062111} (\bibinfo{year}{2013}{\natexlab{a}}),
  \urlprefix\url{https://link.aps.org/doi/10.1103/PhysRevA.88.062111}.

\bibitem[{\citenamefont{Bender et~al.}(2013{\natexlab{b}})\citenamefont{Bender,
  Berntson, Parker, and Samuel}}]{doi:10.1119/1.4789549}
\bibinfo{author}{\bibfnamefont{C.~M.} \bibnamefont{Bender}},
  \bibinfo{author}{\bibfnamefont{B.~K.} \bibnamefont{Berntson}},
  \bibinfo{author}{\bibfnamefont{D.}~\bibnamefont{Parker}}, \bibnamefont{and}
  \bibinfo{author}{\bibfnamefont{E.}~\bibnamefont{Samuel}},
  \bibinfo{journal}{American Journal of Physics} \textbf{\bibinfo{volume}{81}},
  \bibinfo{pages}{173} (\bibinfo{year}{2013}{\natexlab{b}}),
  \eprint{https://doi.org/10.1119/1.4789549},
  \urlprefix\url{https://doi.org/10.1119/1.4789549}.

\bibitem[{\citenamefont{Bender et~al.}(2014)\citenamefont{Bender, Gianfreda,
  and Klevansky}}]{PhysRevA.90.022114}
\bibinfo{author}{\bibfnamefont{C.~M.} \bibnamefont{Bender}},
  \bibinfo{author}{\bibfnamefont{M.}~\bibnamefont{Gianfreda}},
  \bibnamefont{and} \bibinfo{author}{\bibfnamefont{S.~P.}
  \bibnamefont{Klevansky}}, \bibinfo{journal}{Phys. Rev. A}
  \textbf{\bibinfo{volume}{90}}, \bibinfo{pages}{022114}
  (\bibinfo{year}{2014}),
  \urlprefix\url{https://link.aps.org/doi/10.1103/PhysRevA.90.022114}.

\bibitem[{\citenamefont{Chtchelkatchev
  et~al.}(2012)\citenamefont{Chtchelkatchev, Golubov, Baturina, and
  Vinokur}}]{PhysRevLett.109.150405}
\bibinfo{author}{\bibfnamefont{N.~M.} \bibnamefont{Chtchelkatchev}},
  \bibinfo{author}{\bibfnamefont{A.~A.} \bibnamefont{Golubov}},
  \bibinfo{author}{\bibfnamefont{T.~I.} \bibnamefont{Baturina}},
  \bibnamefont{and} \bibinfo{author}{\bibfnamefont{V.~M.}
  \bibnamefont{Vinokur}}, \bibinfo{journal}{Phys. Rev. Lett.}
  \textbf{\bibinfo{volume}{109}}, \bibinfo{pages}{150405}
  (\bibinfo{year}{2012}),
  \urlprefix\url{https://link.aps.org/doi/10.1103/PhysRevLett.109.150405}.

\bibitem[{\citenamefont{Feng et~al.}(2017)\citenamefont{Feng, El-Ganainy, and
  Ge}}]{Feng2017}
\bibinfo{author}{\bibfnamefont{L.}~\bibnamefont{Feng}},
  \bibinfo{author}{\bibfnamefont{R.}~\bibnamefont{El-Ganainy}},
  \bibnamefont{and} \bibinfo{author}{\bibfnamefont{L.}~\bibnamefont{Ge}},
  \bibinfo{journal}{Nature Photonics} \textbf{\bibinfo{volume}{11}},
  \bibinfo{pages}{752} (\bibinfo{year}{2017}),
  \urlprefix\url{https://doi.org/10.1038/s41566-017-0031-1}.

\bibitem[{\citenamefont{Lin et~al.}(2011)\citenamefont{Lin, Ramezani,
  Eichelkraut, Kottos, Cao, and Christodoulides}}]{PhysRevLett.106.213901}
\bibinfo{author}{\bibfnamefont{Z.}~\bibnamefont{Lin}},
  \bibinfo{author}{\bibfnamefont{H.}~\bibnamefont{Ramezani}},
  \bibinfo{author}{\bibfnamefont{T.}~\bibnamefont{Eichelkraut}},
  \bibinfo{author}{\bibfnamefont{T.}~\bibnamefont{Kottos}},
  \bibinfo{author}{\bibfnamefont{H.}~\bibnamefont{Cao}}, \bibnamefont{and}
  \bibinfo{author}{\bibfnamefont{D.~N.} \bibnamefont{Christodoulides}},
  \bibinfo{journal}{Phys. Rev. Lett.} \textbf{\bibinfo{volume}{106}},
  \bibinfo{pages}{213901} (\bibinfo{year}{2011}),
  \urlprefix\url{https://link.aps.org/doi/10.1103/PhysRevLett.106.213901}.

\bibitem[{\citenamefont{Feng et~al.}(2013)\citenamefont{Feng, Xu, Fegadolli,
  Lu, Oliveira, Almeida, Chen, and Scherer}}]{Feng2013}
\bibinfo{author}{\bibfnamefont{L.}~\bibnamefont{Feng}},
  \bibinfo{author}{\bibfnamefont{Y.-L.} \bibnamefont{Xu}},
  \bibinfo{author}{\bibfnamefont{W.~S.} \bibnamefont{Fegadolli}},
  \bibinfo{author}{\bibfnamefont{M.-H.} \bibnamefont{Lu}},
  \bibinfo{author}{\bibfnamefont{J.~B.} \bibnamefont{Oliveira}},
  \bibinfo{author}{\bibfnamefont{V.~R.} \bibnamefont{Almeida}},
  \bibinfo{author}{\bibfnamefont{Y.-F.} \bibnamefont{Chen}}, \bibnamefont{and}
  \bibinfo{author}{\bibfnamefont{A.}~\bibnamefont{Scherer}},
  \bibinfo{journal}{Nature Materials} \textbf{\bibinfo{volume}{12}},
  \bibinfo{pages}{108} (\bibinfo{year}{2013}),
  \urlprefix\url{https://doi.org/10.1038/nmat3495}.

\bibitem[{\citenamefont{Mostafazadeh}(2013)}]{PhysRevA.87.012103}
\bibinfo{author}{\bibfnamefont{A.}~\bibnamefont{Mostafazadeh}},
  \bibinfo{journal}{Phys. Rev. A} \textbf{\bibinfo{volume}{87}},
  \bibinfo{pages}{012103} (\bibinfo{year}{2013}),
  \urlprefix\url{https://link.aps.org/doi/10.1103/PhysRevA.87.012103}.

\bibitem[{\citenamefont{Makris et~al.}(2008)\citenamefont{Makris, El-Ganainy,
  Christodoulides, and Musslimani}}]{PhysRevLett.100.103904}
\bibinfo{author}{\bibfnamefont{K.~G.} \bibnamefont{Makris}},
  \bibinfo{author}{\bibfnamefont{R.}~\bibnamefont{El-Ganainy}},
  \bibinfo{author}{\bibfnamefont{D.~N.} \bibnamefont{Christodoulides}},
  \bibnamefont{and} \bibinfo{author}{\bibfnamefont{Z.~H.}
  \bibnamefont{Musslimani}}, \bibinfo{journal}{Phys. Rev. Lett.}
  \textbf{\bibinfo{volume}{100}}, \bibinfo{pages}{103904}
  (\bibinfo{year}{2008}),
  \urlprefix\url{https://link.aps.org/doi/10.1103/PhysRevLett.100.103904}.

\bibitem[{\citenamefont{Regensburger et~al.}(2013)\citenamefont{Regensburger,
  Miri, Bersch, N\"ager, Onishchukov, Christodoulides, and
  Peschel}}]{PhysRevLett.110.223902}
\bibinfo{author}{\bibfnamefont{A.}~\bibnamefont{Regensburger}},
  \bibinfo{author}{\bibfnamefont{M.-A.} \bibnamefont{Miri}},
  \bibinfo{author}{\bibfnamefont{C.}~\bibnamefont{Bersch}},
  \bibinfo{author}{\bibfnamefont{J.}~\bibnamefont{N\"ager}},
  \bibinfo{author}{\bibfnamefont{G.}~\bibnamefont{Onishchukov}},
  \bibinfo{author}{\bibfnamefont{D.~N.} \bibnamefont{Christodoulides}},
  \bibnamefont{and} \bibinfo{author}{\bibfnamefont{U.}~\bibnamefont{Peschel}},
  \bibinfo{journal}{Phys. Rev. Lett.} \textbf{\bibinfo{volume}{110}},
  \bibinfo{pages}{223902} (\bibinfo{year}{2013}),
  \urlprefix\url{https://link.aps.org/doi/10.1103/PhysRevLett.110.223902}.

\bibitem[{\citenamefont{Garmon et~al.}(2015)\citenamefont{Garmon, Gianfreda,
  and Hatano}}]{PhysRevA.92.022125}
\bibinfo{author}{\bibfnamefont{S.}~\bibnamefont{Garmon}},
  \bibinfo{author}{\bibfnamefont{M.}~\bibnamefont{Gianfreda}},
  \bibnamefont{and} \bibinfo{author}{\bibfnamefont{N.}~\bibnamefont{Hatano}},
  \bibinfo{journal}{Phys. Rev. A} \textbf{\bibinfo{volume}{92}},
  \bibinfo{pages}{022125} (\bibinfo{year}{2015}),
  \urlprefix\url{https://link.aps.org/doi/10.1103/PhysRevA.92.022125}.

\bibitem[{\citenamefont{Guo and
  Gasparian}(2022{\natexlab{a}})}]{PhysRevResearch.4.023083}
\bibinfo{author}{\bibfnamefont{P.}~\bibnamefont{Guo}} \bibnamefont{and}
  \bibinfo{author}{\bibfnamefont{V.}~\bibnamefont{Gasparian}},
  \bibinfo{journal}{Phys. Rev. Research} \textbf{\bibinfo{volume}{4}},
  \bibinfo{pages}{023083} (\bibinfo{year}{2022}{\natexlab{a}}),
  \urlprefix\url{https://link.aps.org/doi/10.1103/PhysRevResearch.4.023083}.

\bibitem[{\citenamefont{Wigner}(1955)}]{PhysRev.98.145}
\bibinfo{author}{\bibfnamefont{E.~P.} \bibnamefont{Wigner}},
  \bibinfo{journal}{Phys. Rev.} \textbf{\bibinfo{volume}{98}},
  \bibinfo{pages}{145} (\bibinfo{year}{1955}),
  \urlprefix\url{https://link.aps.org/doi/10.1103/PhysRev.98.145}.

\bibitem[{\citenamefont{Smith}(1960)}]{PhysRev.118.349}
\bibinfo{author}{\bibfnamefont{F.~T.} \bibnamefont{Smith}},
  \bibinfo{journal}{Phys. Rev.} \textbf{\bibinfo{volume}{118}},
  \bibinfo{pages}{349} (\bibinfo{year}{1960}),
  \urlprefix\url{https://link.aps.org/doi/10.1103/PhysRev.118.349}.

\bibitem[{\citenamefont{Goldberger and Watson}(2004)}]{goldberger2004collision}
\bibinfo{author}{\bibfnamefont{M.}~\bibnamefont{Goldberger}} \bibnamefont{and}
  \bibinfo{author}{\bibfnamefont{K.}~\bibnamefont{Watson}},
  \emph{\bibinfo{title}{Collision Theory}}, Dover books on physics
  (\bibinfo{publisher}{Dover Publications}, \bibinfo{year}{2004}), ISBN
  \bibinfo{isbn}{9780486435077},
  \urlprefix\url{https://books.google.com/books?id=4JUCFZiZOHgC}.

\bibitem[{\citenamefont{Feshbach}(1985)}]{FESHBACH1985398}
\bibinfo{author}{\bibfnamefont{H.}~\bibnamefont{Feshbach}},
  \bibinfo{journal}{Annals of Physics} \textbf{\bibinfo{volume}{165}},
  \bibinfo{pages}{398} (\bibinfo{year}{1985}), ISSN \bibinfo{issn}{0003-4916},
  \urlprefix\url{https://www.sciencedirect.com/science/article/pii/0003491685903033}.

\bibitem[{\citenamefont{Brody}(2013)}]{Brody_2013}
\bibinfo{author}{\bibfnamefont{D.~C.} \bibnamefont{Brody}},
  \bibinfo{journal}{Journal of Physics A: Mathematical and Theoretical}
  \textbf{\bibinfo{volume}{47}}, \bibinfo{pages}{035305}
  (\bibinfo{year}{2013}),
  \urlprefix\url{https://doi.org/10.1088/1751-8113/47/3/035305}.

\bibitem[{\citenamefont{Ruschhaupt et~al.}(2005)\citenamefont{Ruschhaupt,
  Delgado, and Muga}}]{Ruschhaupt_2005}
\bibinfo{author}{\bibfnamefont{A.}~\bibnamefont{Ruschhaupt}},
  \bibinfo{author}{\bibfnamefont{F.}~\bibnamefont{Delgado}}, \bibnamefont{and}
  \bibinfo{author}{\bibfnamefont{J.~G.} \bibnamefont{Muga}},
  \bibinfo{journal}{Journal of Physics A: Mathematical and General}
  \textbf{\bibinfo{volume}{38}}, \bibinfo{pages}{L171} (\bibinfo{year}{2005}),
  \urlprefix\url{https://dx.doi.org/10.1088/0305-4470/38/9/L03}.

\bibitem[{\citenamefont{Gasparian et~al.}(1988)\citenamefont{Gasparian,
  Altshuler, Aronov, and Kasamanian}}]{GASPARIAN1988201}
\bibinfo{author}{\bibfnamefont{V.}~\bibnamefont{Gasparian}},
  \bibinfo{author}{\bibfnamefont{B.}~\bibnamefont{Altshuler}},
  \bibinfo{author}{\bibfnamefont{A.}~\bibnamefont{Aronov}}, \bibnamefont{and}
  \bibinfo{author}{\bibfnamefont{Z.}~\bibnamefont{Kasamanian}},
  \bibinfo{journal}{Physics Letters A} \textbf{\bibinfo{volume}{132}},
  \bibinfo{pages}{201} (\bibinfo{year}{1988}), ISSN \bibinfo{issn}{0375-9601},
  \urlprefix\url{https://www.sciencedirect.com/science/article/pii/0375960188902848}.

\bibitem[{\citenamefont{Gasparian et~al.}(1997)\citenamefont{Gasparian,
  Gummich, Jódar, Ruiz, and Ortuño}}]{GASPARIAN199772}
\bibinfo{author}{\bibfnamefont{V.}~\bibnamefont{Gasparian}},
  \bibinfo{author}{\bibfnamefont{U.}~\bibnamefont{Gummich}},
  \bibinfo{author}{\bibfnamefont{E.}~\bibnamefont{Jódar}},
  \bibinfo{author}{\bibfnamefont{J.}~\bibnamefont{Ruiz}}, \bibnamefont{and}
  \bibinfo{author}{\bibfnamefont{M.}~\bibnamefont{Ortuño}},
  \bibinfo{journal}{Physica B: Condensed Matter}
  \textbf{\bibinfo{volume}{233}}, \bibinfo{pages}{72} (\bibinfo{year}{1997}),
  ISSN \bibinfo{issn}{0921-4526},
  \urlprefix\url{https://www.sciencedirect.com/science/article/pii/S0921452696010368}.

\bibitem[{\citenamefont{Aronov et~al.}(1991)\citenamefont{Aronov, Gasparian,
  and Gummich}}]{Aronov_1991}
\bibinfo{author}{\bibfnamefont{A.~G.} \bibnamefont{Aronov}},
  \bibinfo{author}{\bibfnamefont{V.~M.} \bibnamefont{Gasparian}},
  \bibnamefont{and} \bibinfo{author}{\bibfnamefont{U.}~\bibnamefont{Gummich}},
  \bibinfo{journal}{Journal of Physics: Condensed Matter}
  \textbf{\bibinfo{volume}{3}}, \bibinfo{pages}{3023} (\bibinfo{year}{1991}),
  \urlprefix\url{https://doi.org/10.1088/0953-8984/3/17/017}.

\bibitem[{\citenamefont{Korringa}(1947)}]{KORRINGA1947392}
\bibinfo{author}{\bibfnamefont{J.}~\bibnamefont{Korringa}},
  \bibinfo{journal}{Physica} \textbf{\bibinfo{volume}{13}},
  \bibinfo{pages}{392} (\bibinfo{year}{1947}), ISSN \bibinfo{issn}{0031-8914},
  \urlprefix\url{https://www.sciencedirect.com/science/article/pii/003189144790013X}.

\bibitem[{\citenamefont{Kohn and Rostoker}(1954)}]{PhysRev.94.1111}
\bibinfo{author}{\bibfnamefont{W.}~\bibnamefont{Kohn}} \bibnamefont{and}
  \bibinfo{author}{\bibfnamefont{N.}~\bibnamefont{Rostoker}},
  \bibinfo{journal}{Phys. Rev.} \textbf{\bibinfo{volume}{94}},
  \bibinfo{pages}{1111} (\bibinfo{year}{1954}),
  \urlprefix\url{https://link.aps.org/doi/10.1103/PhysRev.94.1111}.

\bibitem[{\citenamefont{L{\"u}scher}(1991)}]{Luscher:1990ux}
\bibinfo{author}{\bibfnamefont{M.}~\bibnamefont{L{\"u}scher}},
  \bibinfo{journal}{Nucl. Phys.} \textbf{\bibinfo{volume}{B354}},
  \bibinfo{pages}{531} (\bibinfo{year}{1991}).

\bibitem[{\citenamefont{Busch et~al.}(1998)\citenamefont{Busch, Englert,
  Rza\.zewski, and Wilkens}}]{Busch98}
\bibinfo{author}{\bibfnamefont{T.}~\bibnamefont{Busch}},
  \bibinfo{author}{\bibfnamefont{B.-G.} \bibnamefont{Englert}},
  \bibinfo{author}{\bibfnamefont{K.}~\bibnamefont{Rza\.zewski}},
  \bibnamefont{and} \bibinfo{author}{\bibfnamefont{M.}~\bibnamefont{Wilkens}},
  \bibinfo{journal}{Found. Phys.} \textbf{\bibinfo{volume}{28}},
  \bibinfo{pages}{549–559} (\bibinfo{year}{1998}).

\bibitem[{\citenamefont{Guo and Long}(2022)}]{Guo_2022_JPG}
\bibinfo{author}{\bibfnamefont{P.}~\bibnamefont{Guo}} \bibnamefont{and}
  \bibinfo{author}{\bibfnamefont{B.}~\bibnamefont{Long}},
  \bibinfo{journal}{Journal of Physics G: Nuclear and Particle Physics}
  \textbf{\bibinfo{volume}{49}}, \bibinfo{pages}{055104}
  (\bibinfo{year}{2022}),
  \urlprefix\url{https://doi.org/10.1088/1361-6471/ac59d5}.

\bibitem[{\citenamefont{Guo and Gasparian}(2021)}]{PhysRevD.103.094520}
\bibinfo{author}{\bibfnamefont{P.}~\bibnamefont{Guo}} \bibnamefont{and}
  \bibinfo{author}{\bibfnamefont{V.}~\bibnamefont{Gasparian}},
  \bibinfo{journal}{Phys. Rev. D} \textbf{\bibinfo{volume}{103}},
  \bibinfo{pages}{094520} (\bibinfo{year}{2021}),
  \urlprefix\url{https://link.aps.org/doi/10.1103/PhysRevD.103.094520}.

\bibitem[{\citenamefont{Guo and Gasparian}(2022{\natexlab{b}})}]{Guo_2022_JPA}
\bibinfo{author}{\bibfnamefont{P.}~\bibnamefont{Guo}} \bibnamefont{and}
  \bibinfo{author}{\bibfnamefont{V.}~\bibnamefont{Gasparian}},
  \bibinfo{journal}{Journal of Physics A: Mathematical and Theoretical}
  \textbf{\bibinfo{volume}{55}}, \bibinfo{pages}{265201}
  (\bibinfo{year}{2022}{\natexlab{b}}),
  \urlprefix\url{https://doi.org/10.1088/1751-8121/ac7180}.

\bibitem[{\citenamefont{Guo}(2021)}]{PhysRevC.103.064611}
\bibinfo{author}{\bibfnamefont{P.}~\bibnamefont{Guo}}, \bibinfo{journal}{Phys.
  Rev. C} \textbf{\bibinfo{volume}{103}}, \bibinfo{pages}{064611}
  (\bibinfo{year}{2021}),
  \urlprefix\url{https://link.aps.org/doi/10.1103/PhysRevC.103.064611}.

\bibitem[{\citenamefont{El-Ganainy et~al.}(2018)\citenamefont{El-Ganainy,
  Makris, Khajavikhan, Musslimani, Rotter, and
  Christodoulides}}]{10.1038/nphys4323}
\bibinfo{author}{\bibfnamefont{R.}~\bibnamefont{El-Ganainy}},
  \bibinfo{author}{\bibfnamefont{K.~G.} \bibnamefont{Makris}},
  \bibinfo{author}{\bibfnamefont{M.}~\bibnamefont{Khajavikhan}},
  \bibinfo{author}{\bibfnamefont{Z.~H.} \bibnamefont{Musslimani}},
  \bibinfo{author}{\bibfnamefont{S.}~\bibnamefont{Rotter}}, \bibnamefont{and}
  \bibinfo{author}{\bibfnamefont{D.~N.} \bibnamefont{Christodoulides}},
  \bibinfo{journal}{Nature Physics} \textbf{\bibinfo{volume}{14}},
  \bibinfo{pages}{11} (\bibinfo{year}{2018}),
  \urlprefix\url{https://doi.org/10.1038/nphys4323}.

\bibitem[{\citenamefont{Bender et~al.}(2019)\citenamefont{Bender, Dorey,
  Dunning, Fring, Hook, Jones, Kuzhel, Lévai, and Tateo}}]{doi:10.1142/q0178}
\bibinfo{author}{\bibfnamefont{C.~M.} \bibnamefont{Bender}},
  \bibinfo{author}{\bibfnamefont{P.~E.} \bibnamefont{Dorey}},
  \bibinfo{author}{\bibfnamefont{C.}~\bibnamefont{Dunning}},
  \bibinfo{author}{\bibfnamefont{A.}~\bibnamefont{Fring}},
  \bibinfo{author}{\bibfnamefont{D.~W.} \bibnamefont{Hook}},
  \bibinfo{author}{\bibfnamefont{H.~F.} \bibnamefont{Jones}},
  \bibinfo{author}{\bibfnamefont{S.}~\bibnamefont{Kuzhel}},
  \bibinfo{author}{\bibfnamefont{G.}~\bibnamefont{Lévai}}, \bibnamefont{and}
  \bibinfo{author}{\bibfnamefont{R.}~\bibnamefont{Tateo}},
  \emph{\bibinfo{title}{PT Symmetry}} (\bibinfo{publisher}{WORLD SCIENTIFIC
  (EUROPE)}, \bibinfo{year}{2019}),
  \eprint{https://www.worldscientific.com/doi/pdf/10.1142/q0178},
  \urlprefix\url{https://www.worldscientific.com/doi/abs/10.1142/q0178}.

\bibitem[{\citenamefont{Miri and Alù}(2019)}]{doi:10.1126/science.aar7709}
\bibinfo{author}{\bibfnamefont{M.-A.} \bibnamefont{Miri}} \bibnamefont{and}
  \bibinfo{author}{\bibfnamefont{A.}~\bibnamefont{Alù}},
  \bibinfo{journal}{Science} \textbf{\bibinfo{volume}{363}},
  \bibinfo{pages}{eaar7709} (\bibinfo{year}{2019}),
  \eprint{https://www.science.org/doi/pdf/10.1126/science.aar7709},
  \urlprefix\url{https://www.science.org/doi/abs/10.1126/science.aar7709}.

\bibitem[{\citenamefont{{\"O}zdemir et~al.}(2019)\citenamefont{{\"O}zdemir,
  Rotter, Nori, and Yang}}]{doi.org/10.1038/s41563-019-0304-9}
\bibinfo{author}{\bibfnamefont{{\c S}.~K.} \bibnamefont{{\"O}zdemir}},
  \bibinfo{author}{\bibfnamefont{S.}~\bibnamefont{Rotter}},
  \bibinfo{author}{\bibfnamefont{F.}~\bibnamefont{Nori}}, \bibnamefont{and}
  \bibinfo{author}{\bibfnamefont{L.}~\bibnamefont{Yang}},
  \bibinfo{journal}{Nature Materials} \textbf{\bibinfo{volume}{18}},
  \bibinfo{pages}{783} (\bibinfo{year}{2019}),
  \urlprefix\url{https://doi.org/10.1038/s41563-019-0304-9}.

\bibitem[{\citenamefont{Ferise et~al.}(2022)\citenamefont{Ferise, del Hougne,
  F\'elix, Pagneux, and Davy}}]{PhysRevLett.128.203904}
\bibinfo{author}{\bibfnamefont{C.}~\bibnamefont{Ferise}},
  \bibinfo{author}{\bibfnamefont{P.}~\bibnamefont{del Hougne}},
  \bibinfo{author}{\bibfnamefont{S.}~\bibnamefont{F\'elix}},
  \bibinfo{author}{\bibfnamefont{V.}~\bibnamefont{Pagneux}}, \bibnamefont{and}
  \bibinfo{author}{\bibfnamefont{M.}~\bibnamefont{Davy}},
  \bibinfo{journal}{Phys. Rev. Lett.} \textbf{\bibinfo{volume}{128}},
  \bibinfo{pages}{203904} (\bibinfo{year}{2022}),
  \urlprefix\url{https://link.aps.org/doi/10.1103/PhysRevLett.128.203904}.

\bibitem[{\citenamefont{Cutkosky et~al.}(1969)\citenamefont{Cutkosky,
  Landshoff, Olive, and Polkinghorne}}]{CUTKOSKY1969281}
\bibinfo{author}{\bibfnamefont{R.}~\bibnamefont{Cutkosky}},
  \bibinfo{author}{\bibfnamefont{P.}~\bibnamefont{Landshoff}},
  \bibinfo{author}{\bibfnamefont{D.}~\bibnamefont{Olive}}, \bibnamefont{and}
  \bibinfo{author}{\bibfnamefont{J.}~\bibnamefont{Polkinghorne}},
  \bibinfo{journal}{Nuclear Physics B} \textbf{\bibinfo{volume}{12}},
  \bibinfo{pages}{281} (\bibinfo{year}{1969}), ISSN \bibinfo{issn}{0550-3213},
  \urlprefix\url{https://www.sciencedirect.com/science/article/pii/0550321369901692}.

\bibitem[{\citenamefont{Eden et~al.}(2002)\citenamefont{Eden, Landshoff, Olive,
  and Polkinghorne}}]{eden2002analytic}
\bibinfo{author}{\bibfnamefont{R.}~\bibnamefont{Eden}},
  \bibinfo{author}{\bibfnamefont{P.}~\bibnamefont{Landshoff}},
  \bibinfo{author}{\bibfnamefont{D.}~\bibnamefont{Olive}}, \bibnamefont{and}
  \bibinfo{author}{\bibfnamefont{J.}~\bibnamefont{Polkinghorne}},
  \emph{\bibinfo{title}{The Analytic S-Matrix}} (\bibinfo{publisher}{Cambridge
  University Press}, \bibinfo{year}{2002}), ISBN \bibinfo{isbn}{9780521523363},
  \urlprefix\url{https://books.google.com/books?id=VWTnlTyjwjMC}.

\bibitem[{\citenamefont{Mostafazadeh}(2009)}]{PhysRevLett.102.220402}
\bibinfo{author}{\bibfnamefont{A.}~\bibnamefont{Mostafazadeh}},
  \bibinfo{journal}{Phys. Rev. Lett.} \textbf{\bibinfo{volume}{102}},
  \bibinfo{pages}{220402} (\bibinfo{year}{2009}),
  \urlprefix\url{https://link.aps.org/doi/10.1103/PhysRevLett.102.220402}.

\bibitem[{\citenamefont{Ahmed}(2009)}]{Ahmed_2009}
\bibinfo{author}{\bibfnamefont{Z.}~\bibnamefont{Ahmed}},
  \bibinfo{journal}{Journal of Physics A: Mathematical and Theoretical}
  \textbf{\bibinfo{volume}{42}}, \bibinfo{pages}{472005}
  (\bibinfo{year}{2009}),
  \urlprefix\url{https://doi.org/10.1088/1751-8113/42/47/472005}.

\bibitem[{\citenamefont{Longhi}(2009)}]{PhysRevB.80.165125}
\bibinfo{author}{\bibfnamefont{S.}~\bibnamefont{Longhi}},
  \bibinfo{journal}{Phys. Rev. B} \textbf{\bibinfo{volume}{80}},
  \bibinfo{pages}{165125} (\bibinfo{year}{2009}),
  \urlprefix\url{https://link.aps.org/doi/10.1103/PhysRevB.80.165125}.

\bibitem[{\citenamefont{Lloyd}(1969)}]{Lloyd_1969}
\bibinfo{author}{\bibfnamefont{P.}~\bibnamefont{Lloyd}},
  \bibinfo{journal}{Journal of Physics C: Solid State Physics}
  \textbf{\bibinfo{volume}{2}}, \bibinfo{pages}{1717} (\bibinfo{year}{1969}),
  \urlprefix\url{https://doi.org/10.1088/0022-3719/2/10/303}.

\bibitem[{\citenamefont{Ahmed}(2001)}]{AHMED2001231}
\bibinfo{author}{\bibfnamefont{Z.}~\bibnamefont{Ahmed}},
  \bibinfo{journal}{Physics Letters A} \textbf{\bibinfo{volume}{286}},
  \bibinfo{pages}{231} (\bibinfo{year}{2001}), ISSN \bibinfo{issn}{0375-9601},
  \urlprefix\url{https://www.sciencedirect.com/science/article/pii/S0375960101004261}.

\bibitem[{\citenamefont{Guo}(2020)}]{GUO2020135370}
\bibinfo{author}{\bibfnamefont{P.}~\bibnamefont{Guo}},
  \bibinfo{journal}{Physics Letters B} \textbf{\bibinfo{volume}{804}},
  \bibinfo{pages}{135370} (\bibinfo{year}{2020}), ISSN
  \bibinfo{issn}{0370-2693},
  \urlprefix\url{https://www.sciencedirect.com/science/article/pii/S037026932030174X}.

\bibitem[{\citenamefont{Guo}(2013)}]{PhysRevD.88.014507}
\bibinfo{author}{\bibfnamefont{P.}~\bibnamefont{Guo}}, \bibinfo{journal}{Phys.
  Rev. D} \textbf{\bibinfo{volume}{88}}, \bibinfo{pages}{014507}
  (\bibinfo{year}{2013}),
  \urlprefix\url{https://link.aps.org/doi/10.1103/PhysRevD.88.014507}.

\bibitem[{\citenamefont{Guo et~al.}(2013)\citenamefont{Guo, Dudek, Edwards, and
  Szczepaniak}}]{PhysRevD.88.014501}
\bibinfo{author}{\bibfnamefont{P.}~\bibnamefont{Guo}},
  \bibinfo{author}{\bibfnamefont{J.~J.} \bibnamefont{Dudek}},
  \bibinfo{author}{\bibfnamefont{R.~G.} \bibnamefont{Edwards}},
  \bibnamefont{and} \bibinfo{author}{\bibfnamefont{A.~P.}
  \bibnamefont{Szczepaniak}}, \bibinfo{journal}{Phys. Rev. D}
  \textbf{\bibinfo{volume}{88}}, \bibinfo{pages}{014501}
  (\bibinfo{year}{2013}),
  \urlprefix\url{https://link.aps.org/doi/10.1103/PhysRevD.88.014501}.

\bibitem[{\citenamefont{Guo}(2017)}]{PhysRevD.95.054508}
\bibinfo{author}{\bibfnamefont{P.}~\bibnamefont{Guo}}, \bibinfo{journal}{Phys.
  Rev. D} \textbf{\bibinfo{volume}{95}}, \bibinfo{pages}{054508}
  (\bibinfo{year}{2017}),
  \urlprefix\url{https://link.aps.org/doi/10.1103/PhysRevD.95.054508}.

\bibitem[{\citenamefont{Guo and Long}(2020)}]{PhysRevD.101.094510}
\bibinfo{author}{\bibfnamefont{P.}~\bibnamefont{Guo}} \bibnamefont{and}
  \bibinfo{author}{\bibfnamefont{B.}~\bibnamefont{Long}},
  \bibinfo{journal}{Phys. Rev. D} \textbf{\bibinfo{volume}{101}},
  \bibinfo{pages}{094510} (\bibinfo{year}{2020}),
  \urlprefix\url{https://link.aps.org/doi/10.1103/PhysRevD.101.094510}.

\end{thebibliography}

\end{document}